\documentclass{egpubl}
\usepackage{DH2025}%

\usepackage[T1]{fontenc}
\usepackage{dfadobe}  
\usepackage{subcaption}
\usepackage{float}
\usepackage{multirow}
\usepackage{comment}
\usepackage{siunitx}
\usepackage{tablefootnote}
\usepackage{gensymb}
\usepackage[dvipsnames,table]{xcolor}
\usepackage{amsmath}
\usepackage{amsfonts} 
\usepackage{bm}
\usepackage{tikz}

\def\checkmark{\tikz\fill[scale=0.4](0,.35) -- (.25,0) -- (1,.7) -- (.25,.15) -- cycle;}          
\usepackage[ruled,vlined]{algorithm2e}

\usepackage{cite}  
\usepackage{arydshln} 
\usepackage{dblfloatfix} 
\usepackage{hhline} 

\usepackage{eurosym}

\usepackage{booktabs} 

\usepackage[toc,page]{appendix} 

\BibtexOrBiblatex
\electronicVersion
\PrintedOrElectronic

\usepackage{egweblnk} 

\title[Estimating the intrinsic error of 3D point clouds acquisition techniques: A comparative study on instruments of the violin family]%
      {Estimating the intrinsic error of 3D point clouds acquisition techniques: A comparative study on instruments of the violin family}

\author[Philémon Beghin et al.]
{\parbox{\textwidth}{\centering Philémon Beghin$^{1,2}$\orcid{0000-0003-0354-7901}, Jean-Philippe Échard$^{5}$\orcid{0000-0003-1546-6922}, Anthony Peeters$^{1,3}$\orcid{0000-0002-0005-7846}, Arnaud Schenkel$^{6}$\orcid{0009-0001-0632-7114}, \\ Anne-Emmanuelle Ceulemans$^{1,3,4}$\orcid{0000-0002-3273-4004} and François Glineur$^{1,2}$\orcid{0000-0002-5890-1093}
        }
        \\
{\parbox{\textwidth}{\centering $^1$UCLouvain, Louvain-la-Neuve, Belgium\\
         $^2$Institute of Information and Communication Technologies, Electronics and Applied Mathematics (ICTEAM), Louvain-la-Neuve, Belgium\\
         $^3$Institute for the Study of Civilisations, Arts and Letters (INCAL), Louvain-la-Neuve, Belgium\\
         $^4$ Musical Instruments Museum (MIM), Brussels, Belgium \\
         $^5$ Musée de la Musique, Cité de la musique -- Philharmonie de Paris and Centre de Recherche\\ sur la Conservation, CRC-UAR 3224 (CNRS, MNHN, MC) – Sorbonne Université, France \\
         $^6$ PANORAMA, Université Libre de Bruxelles, Brussels, Belgium 
       }
}
}

\begin{document}


\maketitle
\begin{abstract}
This study compares ten digital 3D acquisition techniques applied to violin family instrument plates, assessing their accuracy and relevance for scientific and cultural heritage applications. Evaluated methods include two photogrammetric workflows, medical and industrial computed tomography, terrestrial laser scanning, two structured-light configurations (blue light and infrared), and three consumer-grade iPad Pro approaches: TrueDepth scanning and LiDAR using two different applications. The methods were applied to five historical instruments from the \textit{Musical Instruments Museum} (Brussels) and the \textit{Musée de la Musique} (Paris). Since no geometric ground truth is available, accuracy is assessed through systematic pairwise point-to-plane comparisons of the resulting point clouds. These comparisons are jointly analysed to estimate the intrinsic error of each technique (i.e. the expected discrepancy between the model and the true but unknown geometry) without designating any technique as reference. Results are discussed with respect to intended uses, offering recommendations for techniques depending on the dedicated objectives. 



\end{abstract}  

\section{Introduction}

\noindent In recent years, 3D digitisation through meshes and point clouds has emerged as a valuable tool in the study of heritage objects and particularly musical instruments such as violins \cite{pinto2008photogrammetric,frohlich2009secrets,pyrkosz2011converting,fioravanti2012structural,pyrkosz2013comparative,pyrkosz2014coupled,kirsch2018x,dondi2016measuring,dondi20173d,motteconception,plath2017post,dondi2018multimodal,fuchs2018musices,marschke2018modeling,lothairecharacterization,plath20193d,marschke2020approach,beghindigital,fiocco2021compositional,stanciu2021x,beghin2023validation,meza2025towards}. These instruments, the product of centuries of craftsmanship, are both fragile and unique. 3D technologies provide non-invasive methods to model, document, interpret, and disseminate such artefacts without physical handling or risk of damage. A wide range of applications arise from 3D models, reflecting their growing role across scientific and cultural fields. These include, among others, vibro-acoustic analyses of the violin \cite{pyrkosz2011converting,pyrkosz2013comparative,pyrkosz2014coupled,marschke2018modeling,marschke2020approach,gonzalez2021eigenfrequency,meza2025towards}, deformation and structural study \cite{fioravanti2012structural}, the development of 3D printed affordable instrument replicas \cite{brown2022old,dollins2022design}, a comparison of wood thickness and densities between classical Cremonese and modern instruments \cite{stoel2008comparison,borman2009review}, and studies of the morphological evolution of the violin family, as well as alterations that instruments have undergone over time \cite{motteconception,lothairecharacterization,echard2020bass_violin,echard2020tenor_violin,beghindigital, ceulemans2023baroque, beghin2023validation, beghin2026discussion, beghin2025identification}. Furthermore, 3D models make it possible to simulate the sound of historical instruments that can no longer be played \cite{roda20213d}.

\noindent Acquiring models of musical instruments enables the creation of artificial parametric or CAD models that can be easily digitally modified \cite{knott1987modal,torres2020exploring,meza2025towards} for further experiments, even when not derived from a specific physical specimen \cite{piantadosi2017three, dollins2022design}. Those CAD models allow researchers, for instance, to test how variations in wood thickness affect vibrations, without the need to build multiple nearly identical instruments with slight thickness differences \cite{meza2025towards}.

\noindent Finally, 3D models play a central role in cultural mediation. European cultural institutions, for instance, have set the goal of fully digitising by 2030 their at-risk cultural heritage collections and their most visited cultural and heritage monuments, buildings and sites, underscoring the importance of robust digital models for education, research, and preservation \cite{pletinckx2012europeana,fernie20243d,isaac2024making}. For more information, see the Commission Recommendation (EU) 2021/1970 of 10 November 2021 on a common European data space for cultural heritage\footnote{\url{https://eur-lex.europa.eu/eli/reco/2021/1970/oj}}.

\noindent The diversity in digitisation techniques and their varying levels of resolution, cost, and accessibility raise a fundamental question: which technology is best suited for which purpose, and what level of accuracy can it achieve? This article rigorously compares ten 3D acquisition techniques: photogrammetry (with and without cross-polarisation), medical and industrial computed tomography (CT) scans, terrestrial laser scanning, optical structured-light scanning (using two different modes), and three consumer-grade approaches based on iPad Pro hardware, including TrueDepth and LiDAR scanning using two different applications. We aim to quantitatively assess the intrinsic error associated with each technique. Here, intrinsic error refers to the geometric error attributable to the acquisition technique itself, i.e. the expected discrepancy between the reconstructed geometry and the true geometry of the object. Ideally, this error would be measured directly against an exact reference model. However, no such absolute ground truth is available for the historical instruments considered in this study. The intrinsic errors must therefore be estimated indirectly, by jointly analysing systematic mathematical comparisons between point clouds acquired using the different techniques. We aim to establish a technical and methodological benchmark for researchers and institutions seeking the most appropriate approach to digitising violins or any other small to medium-sized heritage artefacts.

\noindent Beyond the challenge of comparing digitisation techniques, the case study of instruments from the violin family presents particular difficulties for 3D acquisition. The body of these instruments is made of carved and bent wooden pieces, none of which are intended to remain planar in the finished object. Their contours therefore feature very few sharp edges or straight lines. Moreover, the wooden surfaces are covered with organic varnish layers that are partially transparent to visible light. Subsurface light interactions within these coatings can alter the optical response of the surface, potentially generating measurement artefacts and misleading visual impressions of relief. For these reasons, and to ensure that all compared point clouds capture the same geometric information, the analysis is restricted to the violin plates — namely the sound board and the back — rather than full instrument bodies. This choice is motivated by the need to compare homogeneous surface representations across techniques: while optical methods such as photogrammetry capture only the outer envelope of the object, volumetric techniques such as CT scanning also provide internal structural information. Restricting the analysis to the plates ensures that all techniques are evaluated on a common, well-defined surface, and avoids spurious discrepancies arising from differences in the nature of the captured geometry.

\noindent In this paper, we first present the considered acquisition techniques and review the existing literature on their accuracy. We then describe the data acquisition campaigns and the methodology used to derive the point clouds on which the rigorous pairwise comparisons are established. These pairwise comparisons are then jointly analysed to estimate the intrinsic error associated with each acquisition technique, without considering any individual technique as a ground truth. Finally, we report and discuss the results of these point cloud comparisons and the intrinsic error estimates obtained for each technique.

\section{State of the art}

\noindent We first present a brief overview of the physical principles underlying each acquisition technique, highlighting their operating mechanisms and typical applications. Then, we provide a synthesis of quantitative 3D models comparisons reported in the literature, which highlight the relative strengths, accuracies, and limitations of these techniques when applied to heritage objects. Since an independent geometric ground truth is rarely available in this context, most studies rely on pairwise inter-model distances rather than absolute errors, a methodological limitation that the present work aims to address.

\subsection{3D acquisition techniques}

\noindent \textbf{Photogrammetry} is a 3D reconstruction technique based on the processing of multiple overlapping photographs taken from different viewpoints. By detecting and matching common features across images, specialised software applies reconstruction algorithms to generate a dense point cloud, which can then be converted into a mesh and a textured model. This method is highly flexible and low-cost, since it requires only a camera and appropriate lighting conditions, but its accuracy depends strongly on picture quality, surface texture, and acquisition settings. As a result, photogrammetry is widely used in fields such as archaeology \cite{drap2012underwater,mathys2013comparing,chiabrando2013points,khalaf20183d,magnani2020digital,marin2021photogrammetry,fiz2022examples,cavalieri2023so}, architecture \cite{hanke2002architectural,sapirstein2015photogrammetry,tsoraeva2021application}, and cultural heritage documentation \cite{guidi2004high,yilmaz2007importance,hess20103d,chiabrando2015sfm,aicardi2018recent,kingsland2020comparative,farella2022analysing,laurent2025combining}, where portability and accessibility are crucial advantages. Photogrammetry can only cover the outer surface of an object and not its internal structure. Some studies focussing on the violin family acquisition through photogrammetry include \cite{pinto2008photogrammetric,motteconception,patrucco2018image,beghindigital,beghin2023validation,terlindencomparison}.
 
\noindent \textbf{Computed Tomography (CT) scanning} is a volumetric imaging technique that reconstructs three-dimensional models from a series of two-dimensional X-ray projections acquired at multiple rotation angles. The principle relies on differential X-ray attenuation across materials, allowing both external and internal geometries to be captured non-invasively. Originally developed for medical diagnostics, CT scanning has been widely applied to cultural heritage objects \cite{mathys2013comparing}, including musical instruments \cite{stoel2008comparison,borman2009review,frohlich2009secrets,pyrkosz2011converting, pyrkosz2013comparative, pyrkosz2014coupled,lothairecharacterization,beghindigital,stanciu2021x,beghin2023validation,meza2025towards,terlindencomparison}, providing insights into internal structures, assembly methods, and material composition without physical intervention. Typical medical CT systems achieve voxel sizes on the order of a few hundred micrometres, while \textbf{Industrial Computed Tomography ($\mu$-CT)} can reach resolutions down to a few micrometres, enabling the study of fine details such as cracks, varnish layers, or internal wood densities. For the physical study of violins, such accurate scans can be used to reproduce high resolution digital copies of instruments \cite{plath2017post, plath20193d, fuchs2018musices,kirsch2018x}, technical drawings \cite{kirsch2024sarasate} or vibro-acoustical simulation \cite{marschke2018modeling, marschke2020approach}. For museum staff, both medical and industrial CT scanning are highly time consuming, and somehow restrictive. First, all the historical artefacts moved from the museums to the dedicated scanning infrastructures need to be insured. Then, two technologists must be present. The first one adapts the scanner settings to the density and material of the artefacts while the second handles the scanner itself. Moreover, in a medical context, the number of artefacts that can be scanned is limited and the scanners themselves must remain available in case of medical emergency. Indeed, for obvious ethical reasons, patients’ health comes before research. Finally, some instruments carry pathogens and cannot be scanned at all in a medical environment \cite{beghindigital,beghin2023validation}. 

\noindent \textbf{Structured-light} scanning is a 3D acquisition technique that projects a sequence of known light patterns (typically stripes, grids or fringes) onto an object’s surface, and records their deformations with one or more cameras. The geometric displacement of the patterns, analysed through triangulation (measurement of the angle of the deformation), allows highly accurate reconstruction of the object’s surface geometry. This method is particularly well suited for small- to medium-sized objects, as it can achieve sub-millimetre accuracy and capture fine surface details within relatively short acquisition times \cite{fioravanti2012structural,mathys2013comparing,terlindencomparison}. Structured-light acquisitions are carried out quickly, in near real time. However, the technique can face difficulties with highly reflective, translucent, or dark surfaces, which perturb the detection of the projected patterns and reduce measurement reliability. 

\noindent It should be noted that structured-light does not operate on the same principles as a laser scanner (not studied in this paper), although these techniques are similar and sometimes confused in the literature. Structured-light uses the principle of triangulation to reconstruct 3D models, whereas laser scanning can determine distances either by triangulation (angle of the reflected beam), or by time-of-flight (ToF), the travel time of the emitted and reflected light. Furthermore, the projection and the measurement is sequential for the laser scanner (a beam of one dot or one line at a time) while the capture of the structured-light is dense and simultaneous using projected patterns onto larger areas (an entire area in one go). Several violins or cultural heritage artefacts were modelled using laser scanners \cite{hess20103d,mathys2013comparing,dondi2016measuring,dondi20173d,patrucco2018image,fiocco2021compositional}. 

\noindent \textbf{Terrestrial laser scanning (TLS)} is a 3D acquisition technique based on LiDAR (Light Detection and Ranging) technology. It operates by emitting laser pulses - typically in the near-infrared range (e.g. \SI{1064}{\nm}) - toward an object or scene and measuring the time-of-flight or phase shift of the reflected signal, thereby calculating the distance to each sampled point. By systematically sweeping across a scene, the scanner generates dense point clouds that capture large-scale geometries with high accuracy, often achieving millimetre-level over distances ranging from a few metres to several hundred metres. Unlike photogrammetry, which relies on visual features, LiDAR technology (widely speaking) measures distance directly, making it robust under varying lighting conditions and effective in capturing both large-scale topography and fine structural details, depending on the device that is used. Accuracy typically varies depending on the system and range, but it commonly lies within the order of a few millimetres to centimetres. However, since TLS is particularly suited for surveying buildings, landscapes, and archaeological sites \cite{chiabrando2013points,adamopoulos2020critical,cavalieri2023so,laurent2025combining}, where it enables detailed documentation of complex geometries, it is less efficient for small objects due to limited point density and the need for post-processing to handle occlusions and registration between scans \cite{terlindencomparison}.  

\noindent \textbf{LiDAR sensors embedded in consumer devices such as the iPad Pro} operate on the same principle as traditional LiDAR (i.e. measuring the ToF) but at a smaller scale, optimised for portability and everyday use. The iPad Pro LiDAR is designed for short-range scanning (typically up to 5 metres), making it suitable for rapid acquisition of indoor environments or small- to medium-sized objects \cite{vogt2021comparison,abdel2024indoor,terlindencomparison}. 

\noindent \textbf{TrueDepth technology}, introduced in 2017 by Apple in its iPhone and iPad Pro devices, is a depth-sensing system of the structured-light family which combines a dot projector and an infrared (IR) camera. The dot projector emits thousands of infrared points onto the object’s surface, and the IR camera captures their deformation to reconstruct a depth map. While originally designed for biometric facial recognition (Face ID), the system has also been explored for 3D reconstruction and scanning applications. Although less accurate than professional structured-light scanners, its portability, affordability, and integration into consumer devices make it a promising tool for rapid prototyping, medical imaging, and cultural heritage documentation at a low cost \cite{vogt2021comparison,rocuts2024comparison,beltran2025performance,terlindencomparison}.

\subsection{Review of quantitative comparisons between acquisition techniques}

\noindent A number of studies have conducted quantitative comparisons between 3D acquisition techniques, providing valuable insights into their relative accuracy, applicability, and limitations \cite{remondino2011heritage,menna20163d,daneshmand20183d}. It has been shown \cite{adamopoulos2020critical} that, when comparing photogrammetry, structured-light scanners, and a phase-based terrestrial laser scanner on heritage artefacts of different sizes, pairwise Hausdorff distances between photogrammetric models generally remain below 1 mm, while inter-model Hausdorff distances between structured-light and photogrammetric reconstructions are typically in the 1--2 mm range. Scalar field mapping of Hausdorff distances between structured-light scans and TLS models further indicates that most pairwise deviations remain below 2 mm \cite{adamopoulos2020critical}. Similar findings, based on visual inspection of Hausdorff distance maps between structured-light and triangulation laser scanner models, suggest that inter-model deviations are concentrated between 0--0.5 mm, with the structured-light model exhibiting the highest geometric fidelity \cite{graham2017epic}. Structured-light scanners have also been shown to outperform photogrammetry in terms of absolute deviation from a master model reference, with both techniques achieving deviations below 0.1 mm for the majority of the artefact's geometry, although structured-light demonstrated superior overall model quality and proved considerably less sensitive to user expertise and acquisition workflow \cite{gebler2021comparison}. Sub-millimetric pairwise agreement has further been demonstrated when comparing medical photogrammetric reconstructions of bowed string instruments against CT scans used as reference, underlining the complementarity of CT for volumetric information and optical techniques for external morphology \cite{beghin2023validation}. In a broader study where photogrammetry, CT, structured-light, and laser scanning are compared on a human skull (with the structured-light model serving as reference) pairwise deviations measured in the GOM Inspect software show that photogrammetry, CT scanning and triangulation laser scanning achieve a mean deviation of $\pm$0.05 mm, 0.09 mm, and $-$0.06 mm respectively. It was further concluded qualitatively that structured-light yields the most accurate results for external structures, CT scanning is superior for internal structures and reflective materials such as enamel, and photogrammetry offers an excellent balance between portability, price, and quality \cite{mathys2013comparing}. \cite{terlindencomparison} investigated simplified violin models derived from those used in the present study and similarly reported, through pairwise comparisons and an estimation of the intrinsic error based on the comparisons, that $\mu$-CT scanning and structured-light techniques outperformed photogrammetry, which in turn yielded better results than TLS and CT scanning. Except in \cite{terlindencomparison} where no single technique is considered as a reference and in  \cite{gebler2021comparison} which compares reconstructions against a master model of known geometry, the reported values in the aforementioned studies reflect pairwise inter-model distances rather than absolute errors with respect to an independent ground truth. However, several studies implicitly designate one technique as a reference (e.g. structured-light in \cite{mathys2013comparing}, CT scans in \cite{beghin2023validation}, and the laser scanner in \cite{adamopoulos2020critical}). Table \ref{tab:comparison_matrix} provides a summary to synthesize these findings, indicating which techniques have been compared against one another. The accuracy of each of these aforementioned techniques usually lies within the (sub-)millimetre range. 

\noindent Beyond high-end systems, low-cost acquisition techniques have also been assessed. TrueDepth technology has been shown to achieve sub-millimetre pairwise deviations for flatness measurements (mean deviation $\sim$0.16 mm) when compared with a professional structured-light scanner used as reference, although other form tolerances such as straightness and position can exceed 0.5 mm, consistently underperforming relative to industrial scanners \cite{vogt2021comparison}. The iPad LiDAR was found to be not applicable for scanning small objects such as a Lego brick, as the generated mesh was not accurate enough for evaluation \cite{vogt2021comparison}. Regarding geometric reconstruction on regular surfaces, TrueDepth has demonstrated an average absolute inter-plane distance error of approximately 0.1 mm at stand-off distances below 250 mm, as measured against a coordinate measurement machine (CMM) used as ground truth, although point cloud dispersion increases significantly with distance, rendering it unsuitable for industrial-level tasks \cite{beltran2025performance}. Comparable performance between TrueDepth-based smartphone applications and professional structured-light scanners has also been reported for facial feature identification, with an overall percentage error of 5.5\% when the structured-light scanner serves as reference \cite{rocuts2024comparison}. By contrast, iPad LiDAR has proven adequate mainly for large-scale indoor mapping, where linear measurements remain within 1--2 cm of TLS reference values under static acquisition conditions \cite{abdel2024indoor}. TLS itself has been benchmarked against dense photogrammetry in architectural heritage using control points measured by total station as absolute reference, revealing centimetric agreement (around 2--3 cm, with the best software achieving below 1 cm) \cite{chiabrando2013points}. Additional works have confirmed that structured-light scanner accuracy \cite{polo2019study} and photogrammetric results \cite{laurent2025combining} are influenced not only by sensor technology, but also by acquisition resolution, software processing, and calibration strategy. 
  
\noindent Taken together, these works confirm that no single technique is universally superior and that the choice of technique must depend above all on the objectives and desired results \cite{adamopoulos2020critical, mathys2013comparing}. Structured-light remains the most reliable choice for external geometries at the sub-millimetre scale \cite{adamopoulos2020critical, mathys2013comparing, gebler2021comparison}, photogrammetry offers a cost-effective and flexible alternative with slightly lower accuracy \cite{adamopoulos2020critical, mathys2013comparing, laurent2025combining}, CT scanning is essential for internal structures \cite{mathys2013comparing, beghin2023validation}, TLS excels for architectural and large-scale contexts \cite{chiabrando2013points, adamopoulos2020critical}, and low-cost sensors provide practical though less accurate options \cite{vogt2021comparison, abdel2024indoor, rocuts2024comparison}. Some studies even suggest hybrid workflows where two or even more techniques are combined together to increase the model quality \cite{guidi2004high,fioravanti2012structural,mathys2013comparing,fiocco2021compositional,dondi2016measuring,dondi20173d,laurent2025combining}. These considerations and recommended uses reflect the conclusions commonly reported in the existing literature. In the following sections, we assess whether similar conclusions and orders of magnitude of accuracy are obtained in the specific context of our study, which focuses on the 3D geometry of violin sound board and back plates. 

\begin{table*}
\centering
\resizebox{\textwidth}{!}{
\begin{tabular}{|l|l|l|l|l|l|l|}
\hline
\textbf{Technique} & \textbf{Medical CT} & \textbf{Industrial $\mu$-CT} & \textbf{Structured-light} & \textbf{TLS} & \textbf{LiDAR (iPad)} & \textbf{TrueDepth} \\
\hline
\textbf{Photogrammetry} & \cite{beghindigital,beghin2023validation,mathys2013comparing,terlindencomparison} &  \cite{terlindencomparison} & \cite{mathys2013comparing,menna20163d,adamopoulos2020critical, graham2017epic,gebler2021comparison,patrucco2018image,terlindencomparison} &\cite{chiabrando2013points,adamopoulos2020critical, laurent2025combining,terlindencomparison} &  &  \\
\hline
\textbf{Medical CT scan} & \cellcolor{lightgray} &  & \cite{mathys2013comparing} & \cite{terlindencomparison} &  &   \\
\hline
\textbf{Industrial $\mu$-CT scan} & \cellcolor{lightgray} & \cellcolor{lightgray} & \cite{terlindencomparison} &  &  &  \\
\hline
\textbf{Structured-light} & \cellcolor{lightgray} & \cellcolor{lightgray} & \cellcolor{lightgray} & \cite{adamopoulos2020critical} & \cite{vogt2021comparison} & \cite{vogt2021comparison,rocuts2024comparison} \\
\hline
\textbf{TLS} & \cellcolor{lightgray} & \cellcolor{lightgray} & \cellcolor{lightgray} & \cellcolor{lightgray} & \cite{abdel2024indoor} &  \\
\hline
\textbf{LiDAR (iPad Pro)} & \cellcolor{lightgray} & \cellcolor{lightgray} & \cellcolor{lightgray} & \cellcolor{lightgray} & \cellcolor{lightgray} &  \\
\hline
\end{tabular}
} 
\caption{Quantified pairwise comparisons between acquisitions techniques found in the literature.}
\label{tab:comparison_matrix}
\end{table*}

\section{Acquisition methodology}

\subsection{Musical Instruments}

\noindent In this study, five instruments from the violin family were digitised by different subsets of the above-described 3D acquisition technologies. Two of these, a viola (inv. no. \href{https://collectionsdumusee.philharmoniedeparis.fr/collectionsdumusee/doc/MUSEE/0157875/tenor-de-violon-recoupe-en-alto}{E.1731}) and a cello (inv. no. \href{https://collectionsdumusee.philharmoniedeparis.fr/collectionsdumusee/doc/MUSEE/1115940/basse-de-violon-recoupee-en-violoncelle}{E.2020.1.1}), were crafted by Andrea Amati and are part of the French national collection (Musée de la Musique, Paris). The remaining three instruments belong to the Musical Instruments Museum in Brussels: a cello by Gaspar Borbon (inv. no. \href{https://www.carmentis.be:443/eMP/eMuseumPlus?service=ExternalInterface&module=collection&objectId=109450&viewType=detailView}{1374}), a cello by Benoît Joseph Boussu (inv. no. \href{https://www.carmentis.be:443/eMP/eMuseumPlus?service=ExternalInterface&module=collection&objectId=109956&viewType=detailView}{2863}), and a violin by Matthys Hofmans IV (inv. no. \href{https://www.carmentis.be:443/eMP/eMuseumPlus?service=ExternalInterface&module=collection&objectId=109427&viewType=detailView}{1355}). For clarity and consistency, the instruments will hereafter be referred to by their inventory numbers (the instruments from Paris will be referred to as 1731 and 2020). 

\subsection{3D acquisition campaigns}
\label{section:3D acquisition campaigns}

\noindent In this section, we present the specifications and acquisition settings of the devices used to generate the 3D point clouds of the musical instruments. Since these technologies produce heterogeneous outputs, all files were subsequently converted into the PLY file format for the sake of standardisation (where `PLY' stands for `Polygon'). The PLY format can encode 3D geometry through vertex coordinates and, when applicable, edges between vertices defining polygonal faces. With the exception of TLS, which directly produces a point cloud, the considered acquisition techniques generate polygonal meshes, from which the sound board and back of each instrument are first delineated automatically following the procedure described in \cite{beghin2023validation}. Since this automatic delineation relies on mesh structure, it cannot be applied to TLS data; the plates were therefore delineated manually in this case. For the systematic comparisons presented in the following sections, only the vertex coordinates are retained across all techniques, ensuring a common geometric representation independently of whether the original output consisted of a mesh or a point cloud. The different technologies and their acquisition settings are first detailed, followed by a summary table indicating which instrument was digitised with each acquisition method. Table \ref{tab:point cloud sizes} in Appendix \ref{section:appendix} provides the number of vertices of each point cloud used in this research.

\begin{figure*}[p]
    \centering
    \setlength{\tabcolsep}{3pt}
    \renewcommand{\arraystretch}{1.0}

    \begin{subfigure}[t]{0.49\textwidth}
        \centering
        \includegraphics[width=0.7\linewidth]{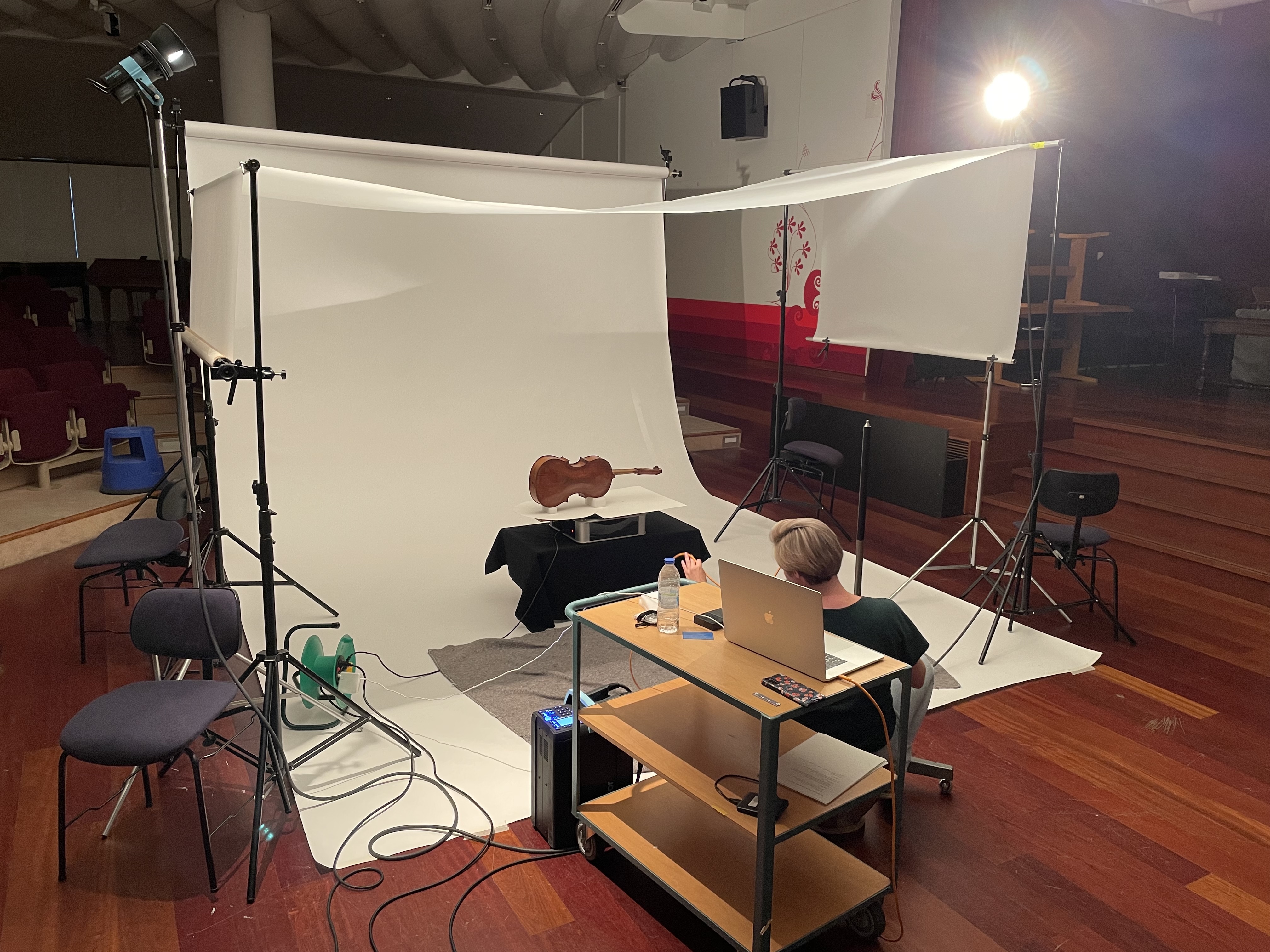}
        \caption{Photogrammetry (Brussels, June 2022)}
        \label{fig:comp-a}
    \end{subfigure}\hfill
    \begin{subfigure}[t]{0.49\textwidth}
        \centering
        \includegraphics[width=0.7\linewidth]{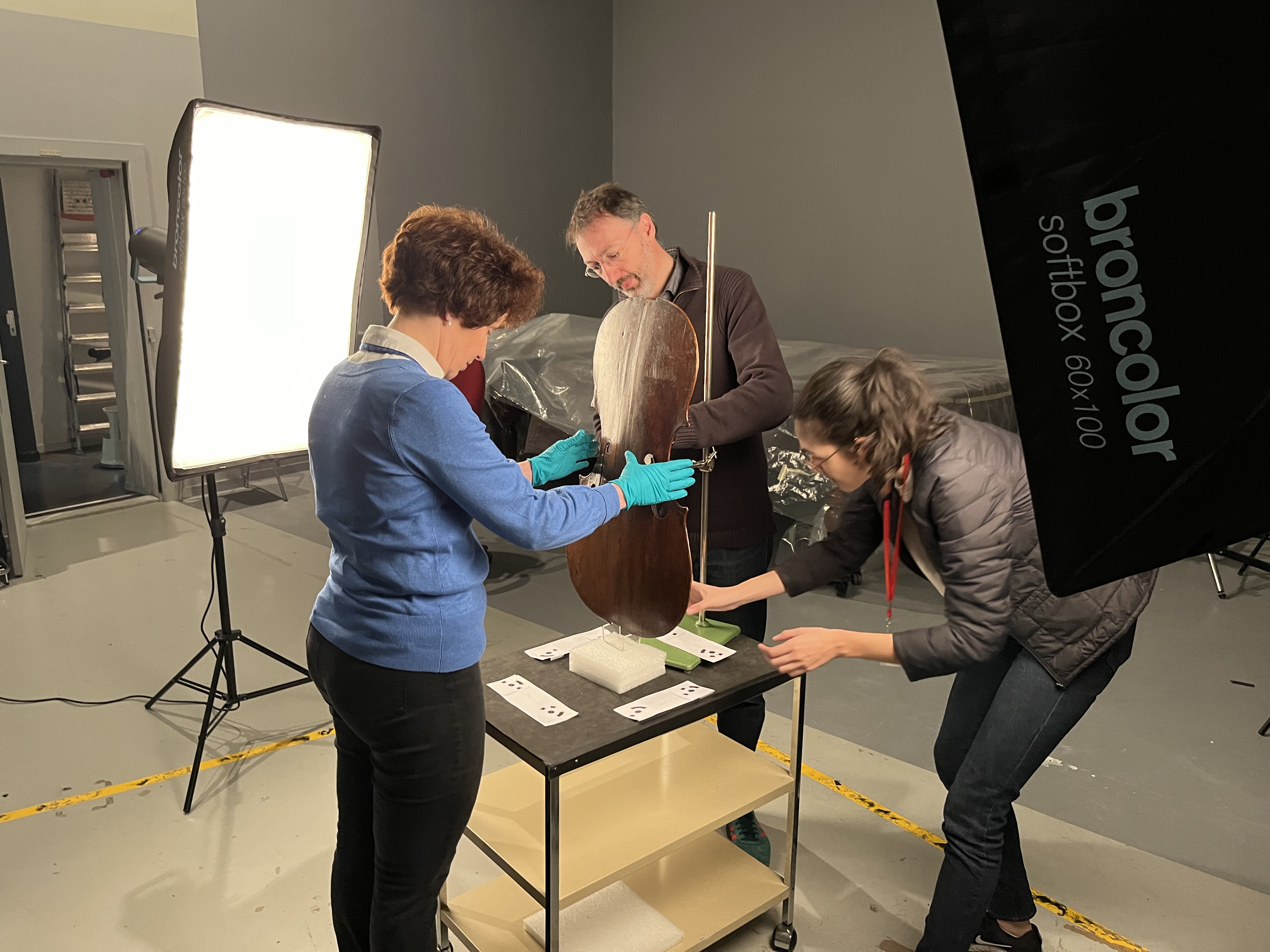}
        \caption{Photogrammetry (Paris, April 2024)}
        \label{fig:comp-b}
    \end{subfigure}

    \vspace{4pt}

    \begin{subfigure}[t]{0.49\textwidth}
        \centering
    \includegraphics[angle=-90,origin=c,height=0.2\textheight]{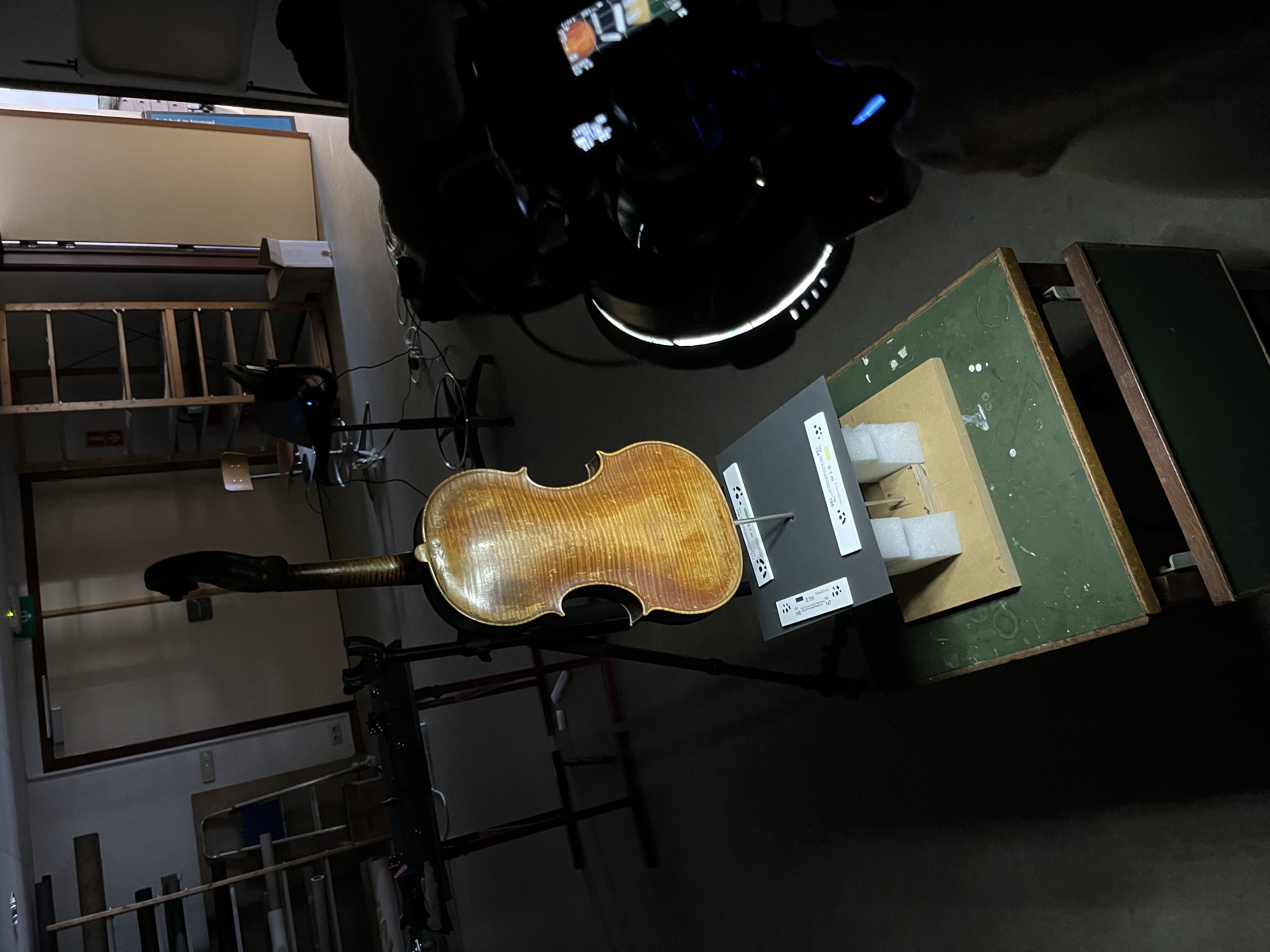}
        \caption{Photogrammetry - PANORAMA (Brussels, October 2025)}
        \label{fig:comp-c}
    \end{subfigure}\hfill
    \begin{subfigure}[t]{0.49\textwidth}
        \centering
    \includegraphics[angle=-90,origin=c,height=0.2\textheight]{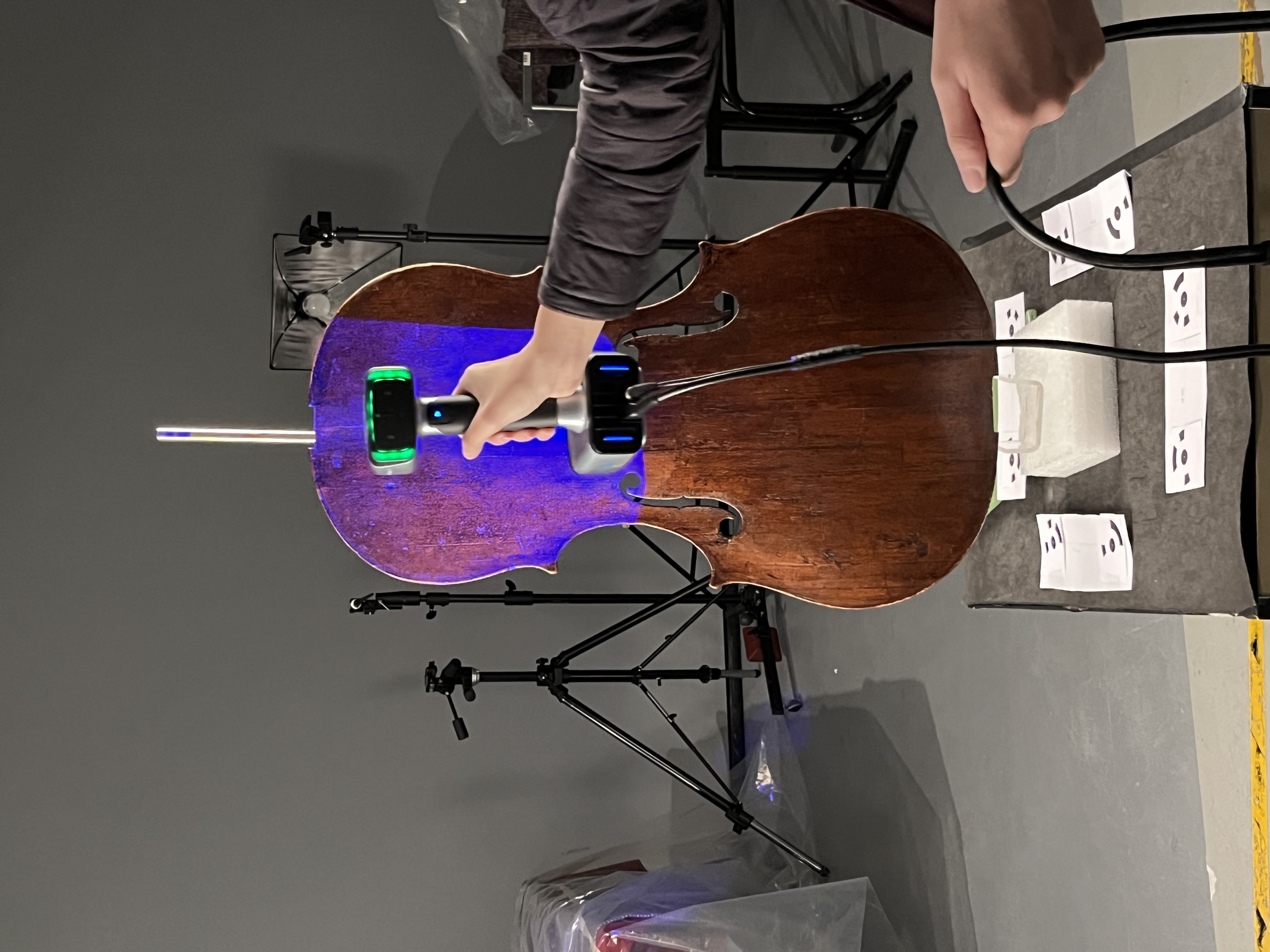}
        \caption{Structured-light - Blue light (Paris, April 2024)}
        \label{fig:comp-d}
    \end{subfigure}

    \vspace{4pt}

    \begin{subfigure}[t]{0.49\textwidth}
        \centering
        \includegraphics[width=0.7\linewidth]{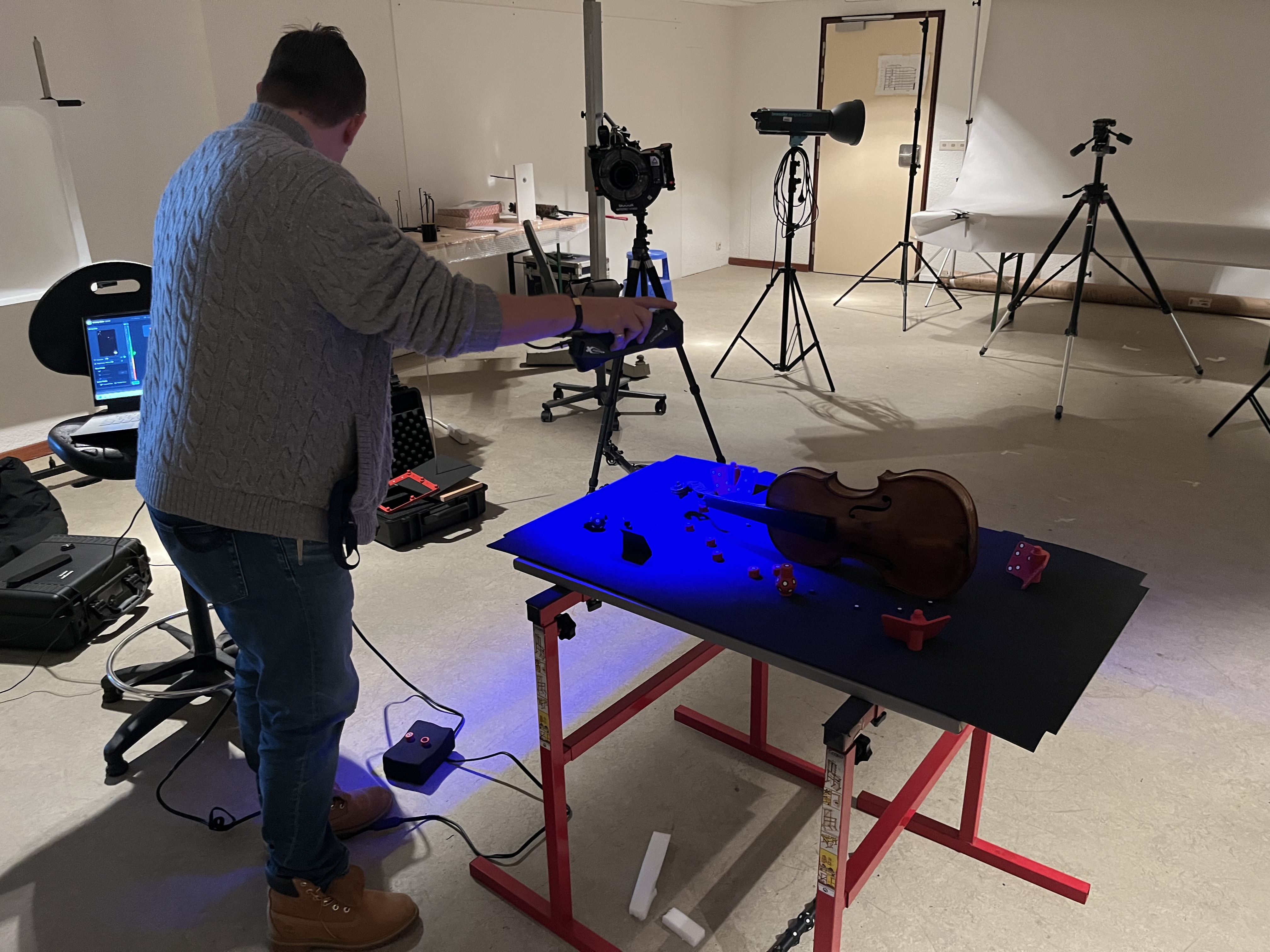}
        \caption{Structured-light - Blue light (aborted) (Brussels, October 2025)}
        \label{fig:comp-e}
    \end{subfigure}\hfill
    \begin{subfigure}[t]{0.49\textwidth}
        \centering
        \includegraphics[width=0.7\linewidth]{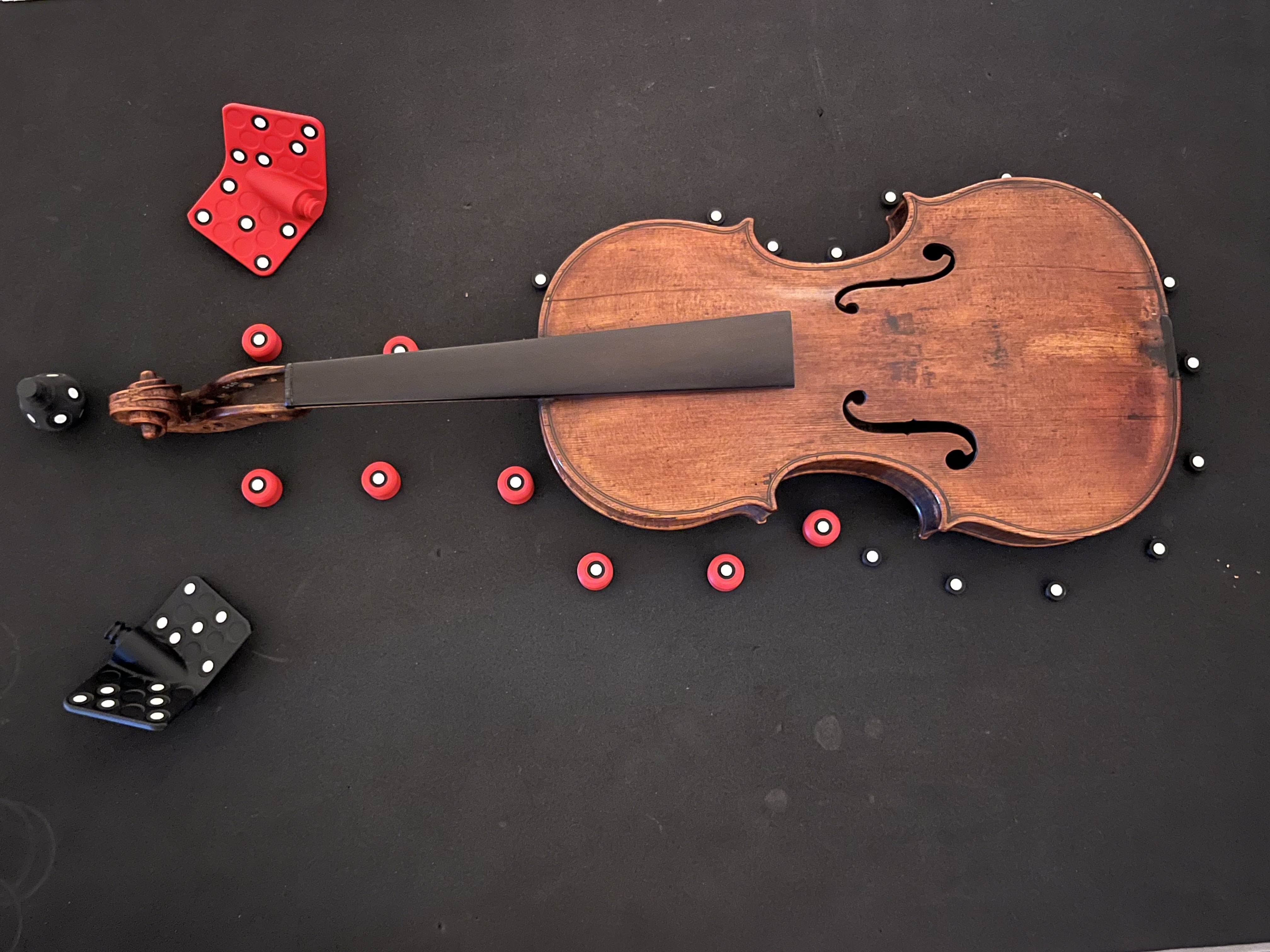}
        \caption{Structured-light - Infrared (Brussels, October 2025)}
        \label{fig:comp-f}
    \end{subfigure}

    \vspace{4pt}

    \begin{subfigure}[t]{0.49\textwidth}
        \centering
        \includegraphics[width=0.7\linewidth]{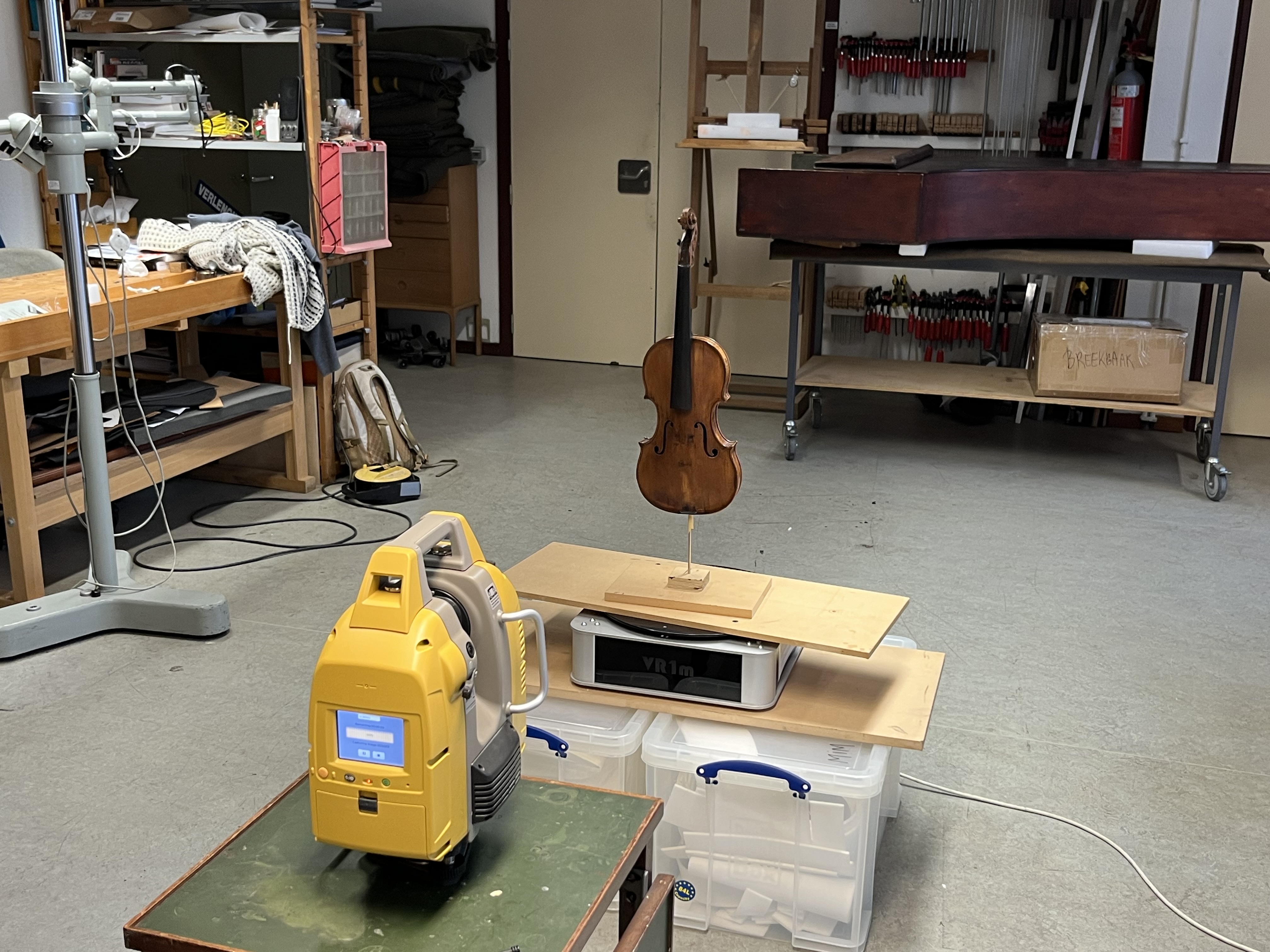}
        \caption{Terrestrial laser scanner (Brussels, April 2024)}
        \label{fig:comp-g}
    \end{subfigure}\hfill
    \begin{subfigure}[t]{0.49\textwidth}
        \centering
        \includegraphics[width=0.7\linewidth]{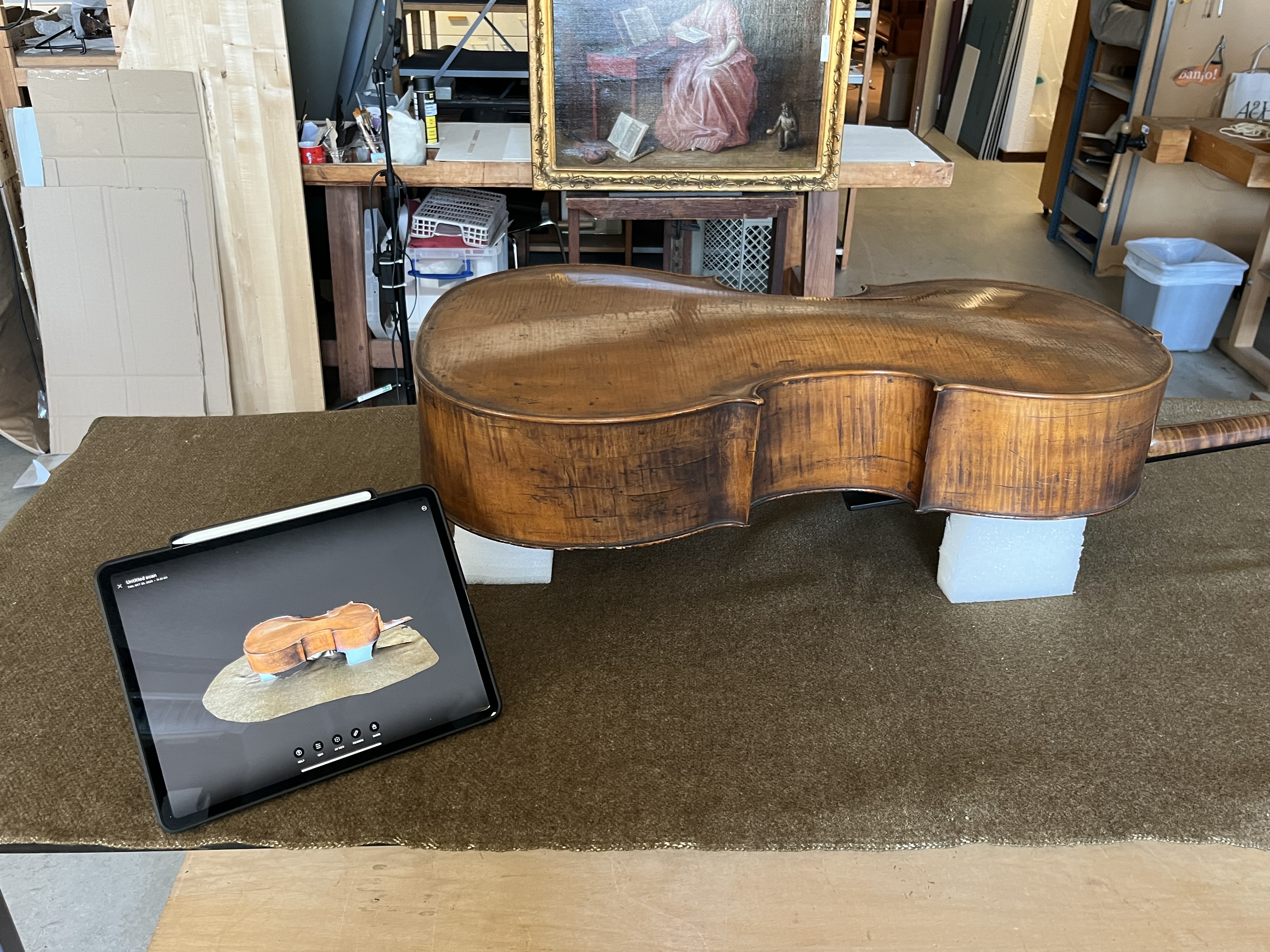}
        \caption{iPad LiDAR - Scaniverse (Brussels, October 2025)}
        \label{fig:comp-h}
    \end{subfigure}
    \caption{Overview of selected acquisition techniques, indicating the year and location of each acquisition}
    \label{fig:composite-8}
\end{figure*}

\begin{itemize}
    \item \textbf{Photogrammetry:} Three photogrammetry acquisition campaigns were carried out. Although they slightly differ from each other (notably due to the available equipment, working practices, and the time gap between them), the first two campaigns are described together, as they still relied on comparable experimental setups and acquisition protocols. The first took place at the Musical Instruments Museum (MIM) in 2022 and included instruments 1355 and 1374, while the second was conducted at the Cité de la Musique in 2024, focusing on the two Andrea Amati instruments (1731 and 2020). \\
    
    \noindent In Brussels, the instruments were positioned on an automatic turntable, rotated through $\ang{360}$, and photographed every $\ang{10}$ from three different heights, in order to cover the whole instruments. The setup, including indirect lighting, is illustrated in Figure \ref{fig:comp-a}. The camera used was a Nikon D850 equipped with a \SI{60}{\mm} focal lens. Each image measured 8256 × 5504 pixels ($\approx$ 50--60 \SI{}{\micro\metre} per pixel) and had an average size of 20 MB. In Paris, where no turntable was available, the instruments were photographed by moving around them, with the photographer maintaining a constant height across successive shots. Figure \ref{fig:comp-b} shows the top plate of instrument 2020 being stabilised. A Nikon D850 was also used, this time with a \SI{50}{\mm} lens. The pixel size was identical and each photo was approximately 5 MB in size. In both campaigns, scale bars were included to ensure direct scaling of the 3D models. The acquisitions were conducted by experienced photographers from the institutions already familiar with photogrammetry: Iona Thys (Brussels) and Julie Toupance (Paris). \\

\noindent Following the acquisition, images were automatically masked in Photoshop to isolate the instruments from their backgrounds, with all masks being systematically checked and corrected when necessary. For each instrument, approximately 150 images and their corresponding masks were imported into Agisoft Metashape. The processing pipeline produced a 3D mesh, which was exported in OBJ format. This widely used format is compatible with most 3D visualisation and editing software, and has been readily converted into PLY format using MeshLab\footnote{\url{https://www.meshlab.net/}} when required. \\

\noindent The third campaign was performed in 2025 at the MIM on violin 1355 and cello 2863 by an independent research platform (PANORAMA, from Université Libre de Bruxelles). The acquisition protocol follows a similar photogrammetric approach than that of the previous two, but uses cross-polarisation to reduce specular reflections, as in Figure \ref{fig:comp-c}. Photographs were taken in a darkened environment using a Nikon Z7 and a ring flash, both fitted with polarising filters \cite{frost2023digital,schenkel2025photogrammetrie,schenkel2025preserving}. Objects are documented with full \ang{360} coverage at multiple heights with strong overlap ($\sim$600 pictures per object), including scale bars for accurate metric referencing. The photogrammetric workflow is processed in RealityCapture: a draft reconstruction is first generated and used to create image masks isolating the object from the background, before final alignment and high-resolution 3D model production. \\

    \item \textbf{CT scanner:} Instruments 1355 and 2863 were both scanned in 2017 at the Saint-Luc University Hospital (UCLouvain, Brussels-Woluwe). The scanner was an IQon - Spectral CT by Philips and it produced 512 $\times$ 512 pixels slices with an overlap rate of 50\%. Each slice was \SI{0.9}{\mm} thick and the size of the pixels is \SI{0.61}{\mm} $\times$ \SI{0.61}{\mm} for the violin 1355 and \SI{0.97}{\mm} $\times$ \SI{0.97}{\mm} for the cello 2863, resulting in overall voxels resolutions of respectively 0.61 $\times$ 0.61 $\times$ \SI{0.45}{\mm} and 0.97 $\times$ 0.97 $\times$ \SI{0.45}{\mm}. The resulting DICOM files (Digital Imaging and Communications in Medicine) were converted into triangular meshes using the 3D Slicer software\footnote{\url{https://www.slicer.org/}, version 5.6.1} \cite{fedorov20123d}, itself using a \textit{Flying Edges} algorithm for the image segmentation \cite{schroeder2015flying}. DICOM is an international standard for storing, exchanging, and transmitting medical images (such as CT, MRI, or X-ray), where the data are typically stored as slices representing cross-sectional views of the body. \\
    
    \item \textbf{$\mu$-CT scanner:} Viola 1731 was $\mu$-CT scanned within the AST-RX platform (UAR 2047 DoHNEE CNRS-MNHN, Paris) in February-March 2022 by Marta Bellato and Patricia Wils. The scanner model is a v|tome|x 240 L from Baker Hughes Digital Solutions. The voxel size is approximately \SI{130}{\micro\m}. The costs for purchasing, running and maintaining such a device are at least greater than the other pieces of equipment here mentioned by one or more orders of magnitude. \\
    
    \item \textbf{Structured-light scanner:} Cello 1374 was digitised in 2023 at the Musical Instruments Museum (MIM), while the two Andrea Amati instruments (1731 and 2020) were scanned in 2024 at the Musée de la Musique. The optical acquisition was performed using the same EinScan HX (Shining 3D), a structured-light scanner with real-time reconstruction of the specimen, shown in Figure \ref{fig:comp-d}. Such a scanner costs about 8000--10000 \officialeuro. The scans were carried out using the blue-light (BL) LED source in Rapid Scan mode, achieving a stated accuracy of \SI{0.3}{\mm}. The output consisted of correctly scaled meshes exported in STL format (STereoLithography). Similar to the OBJ format generated through photogrammetry, STL is widely used and can be readily converted into PLY format using MeshLab. \\

    In addition, a further acquisition campaign was conducted in 2025 at the MIM by the PANORAMA research platform, during which instruments 1374, 2863, and 1355 were digitised. This campaign employed a Creality Raptor X structured-light scanner, as shown in Figure \ref{fig:comp-e}, which costs around 5000 \officialeuro. While the EinScan HX used in Brussels and Paris relied on blue-light projection, this mode proved unsuitable for the PANORAMA acquisitions due to a combination of surface varnish properties and unfavourable light incidence angles on certain regions of the instruments, particularly at the junctions between the ribs and the sound board or back, which are the top and bottom plates of the violin body. To mitigate these effects, PANORAMA acquisitions were performed using the infrared (IR) structured-light mode of the Raptor X scanner. In addition, reflective markers were used, as those in red and black in Figure \ref{fig:comp-f}. These are essential when using a handheld scanner like the Creality Raptor X, as the system relies on tracking the unique and irregular pattern of these markers rather than just the surface geometry, ensuring stable alignment even on shiny, dark, or low-feature objects. For accurate tracking, the markers must be placed randomly with at least four markers visible to the scanner at all times, and with higher density around edges and complex features. As museum objects cannot be directly marked or altered, dedicated 3D-printed support structures are used to hold reflective markers in place. \\
    
    \noindent Although infrared mode provided a better reconstruction of the junction between the sound board and the ribs, the model around this area was still not perfect. A `fold' appeared in certain areas of the edge, creating a second artificial contour to the instrument, as in the planar cut in Figure \ref{fig:SL_problem}. This spurious layer was first removed manually before automatically delineating the plates according to the procedure described in \cite{beghin2023validation}. \\

    \noindent The achievable spatial resolution and measurement accuracy in structured-light systems are influenced by the wavelength of the projected illumination. Shorter wavelengths enable finer spatial sampling of projected patterns, leading to improved detail capture and reduced pattern distortion in triangulation-based reconstruction. Commercial metrology sources\footnote{\url{https://www.keyence.com/products/3d-measure/3d-scanner/resources/3d-scanner-resources/blue-light-vs-white-light-scanner.jsp}} further indicate that blue-light-based structured-light scanners typically achieve higher accuracy and surface detail resolution compared to systems using broad spectrum or longer wavelength illumination. 

    \item \textbf{Terrestrial laser scanner:} Two instruments from the MIM (1355 and 2863) were digitised in Brussels in 2024 using a GLS-2200 (Topcon), a time-of-flight TLS which costs around 80 000 \officialeuro. The device was mounted on a small table, with an optional box to elevate it by \SI{20}{\cm} when scanning the cello. The scanner was positioned at a distance of approximately \SI{2}{\m} from the instrument, as illustrated in Figure \ref{fig:comp-g}. Each scanning session produced a CLR file (Topcon proprietary file format containing a 3D point cloud), first converted into LAS (LASer) file using Magnet Collage software, which could subsequently be converted into other point cloud formats, such as PLY, using MeshLab. This is the only technique which provides only a point cloud and not a mesh. All clouds are already scaled. \\

    Two distinct issues were encountered with the TLS acquisitions. First, reconstruction artefacts were observed inside the f-holes and in the vicinity of the sound board, as in Figure \ref{fig:artefacts_tls}. We manually removed such artefacts using MeshLab. A second issue concerned the structure of the generated point cloud, which exhibited an apparent `quadruple' thickness which we can observe still in Figure \ref{fig:artefacts_tls}. When the acquisition was performed with the laser oriented perpendicular to the sound board or back, the resulting point cloud appeared to consist of four nearly superimposed layers of points, regularly spaced at approximately one millimetre intervals. At present, the precise origin of those two phenomena remains unclear. Potential contributing factors include the varnish properties of the instruments or specific lighting conditions, although the latter should theoretically have limited influence given the time-of-flight operating principle of the scanner. Similar behaviour has also been observed in TLS datasets acquired for other research projects, suggesting that this issue may not be specific to the present study. \\

    To address this second problem already encountered in \cite{terlindencomparison}, a custom algorithm was applied to discretise the point cloud into a regular grid of cells of size $\Delta x \times \Delta y$ \SI{}{\mm} and to retain, within each cell, only the point with the maximum $z$-value. These points are shown in purple in Figure \ref{fig:algo_TLS}, which illustrates this principle with larger grid cells than in the actual calculations for better readability. The domain of the violin plate is shown in brown for information. It is worth noting from Figure \ref{fig:algo_TLS} that, owing to the geometry of the violin plates — whose elevation decreases away from the axis of symmetry — most retained points tend to be located near the edges of the grid cells rather than near their centres. In order to remove the `quadruple' thickness of the TLS point cloud, the square grid cells were progressively increased by setting equally $\Delta x = \Delta y$ until the number of remaining points was reduced to approximately one quarter of the original point count. Although this customised algorithm might retain unwanted points and remove points actually belonging to the violin plate, it remains effective at filtering most of the unwanted vertices of the TLS point cloud. \\

    \begin{figure}
        \centering
        \includegraphics[width=\linewidth]{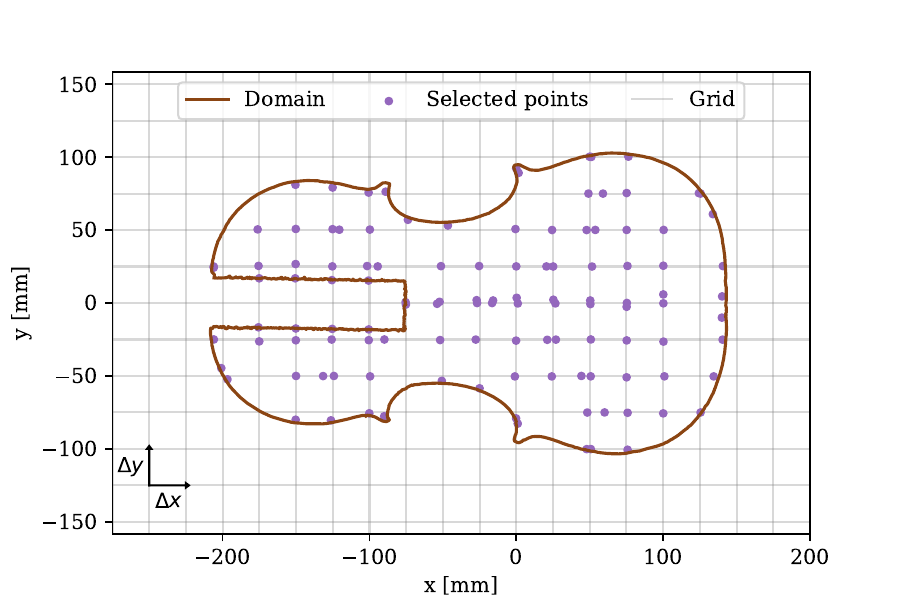}
        \caption{Illustration of the grid-based filtering algorithm used to reduce the TLS point cloud to a single surface layer. The point cloud is discretised into a regular grid of cells of size $\Delta x \times \Delta y$, and only the point with the maximum $z$-value is retained per cell (purple). Grid cells are shown larger than in the actual calculations for readability. The domain of the violin plate is shown in brown as additional information.}
        \label{fig:algo_TLS}
    \end{figure}
    
    \item \textbf{LiDAR and TrueDepth:} Both acquisition technologies were tested in 2025 using the same iPad Pro (4th generation, iPadOS version 17.4.1) whose cost is around 1000 \officialeuro. The three instruments scanned with this device were the violin 1355 and the cellos 1374 and 2863. They were acquired from different angles: either standing on a wooden stick or lying on a table to acquire the sound board and the back separately. Two applications were evaluated : Scaniverse, which supports only LiDAR technology, and Heges, which supports both LiDAR and TrueDepth. In both cases, the scanning procedure consisted of moving around the object (either standing up or laying down) at different heights and angles to capture it. The resulting meshes were directly exported in PLY format, already automatically scaled. Figure \ref{fig:comp-h} illustrates the back of cello 1374 and the 3D model generated with LiDAR technology and the application Scaniverse displayed on the iPad.  
\end{itemize}

\begin{figure}
    \centering
    \includegraphics[width=\linewidth]{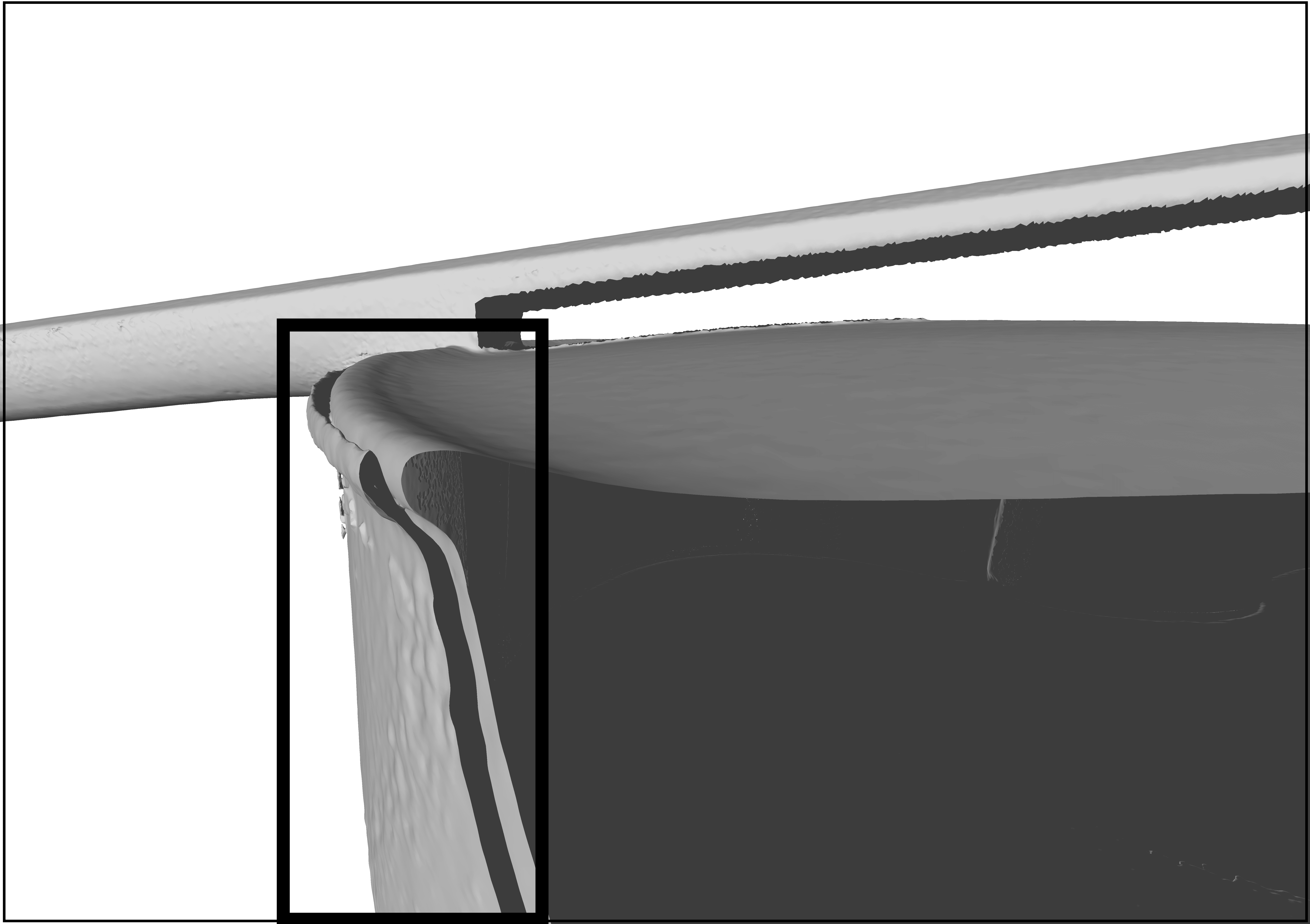}
    \caption{Spurious layer artefact in the structured-light (infrared mode) acquisition of instrument 2863}
    \label{fig:SL_problem}
\end{figure}

\begin{figure}
    \centering
    \includegraphics[width=\linewidth]{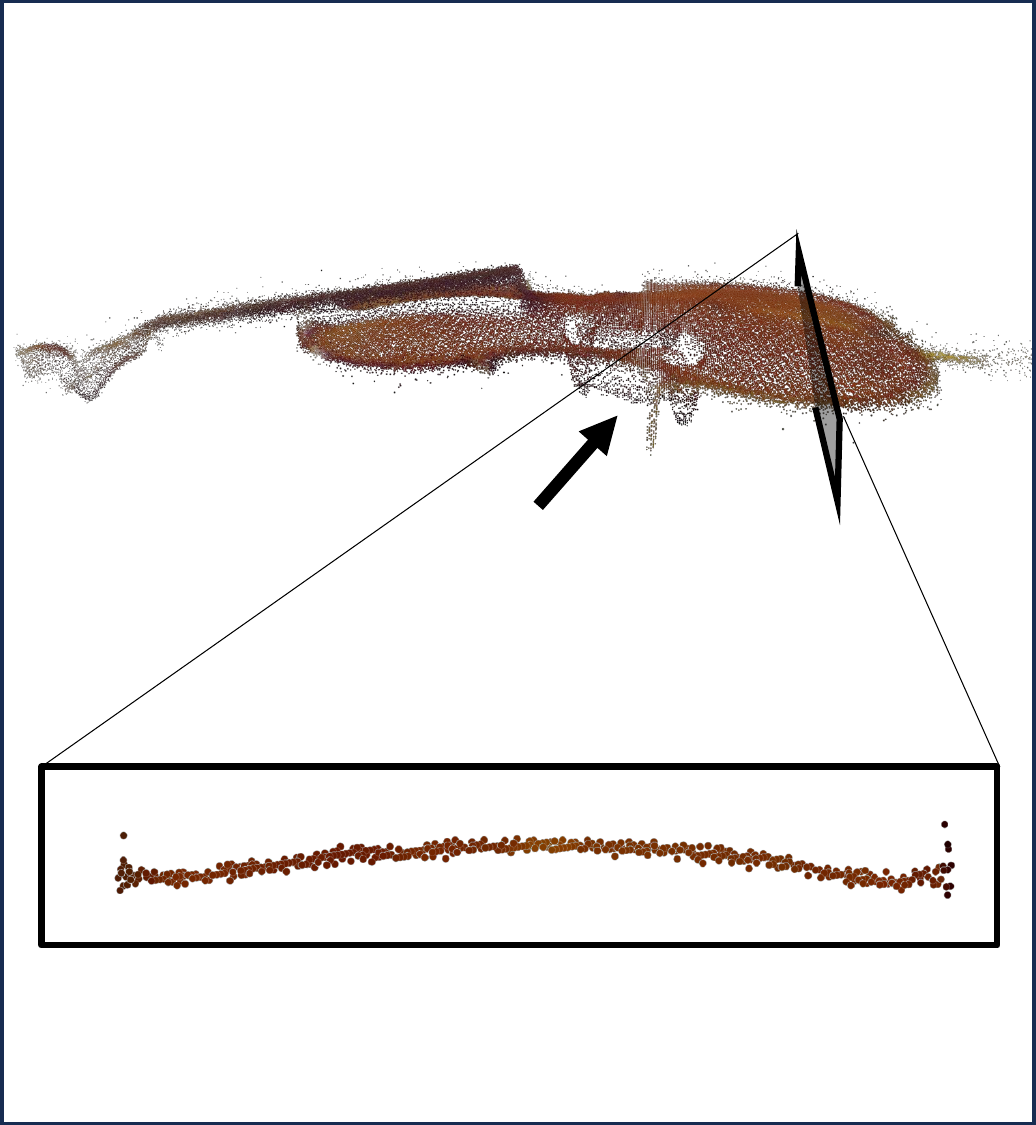}
    \caption{Artefacts in the terrestrial laser scanner acquisitions (indicated in the f-holes with an arrow, and boxed to show the "quadruple" thickness)}
    \label{fig:artefacts_tls}
\end{figure}

\section{Comparison methods}
\label{section:comparison_methods}

\noindent Two major differences must be highlighted among the acquisition techniques considered in this study. First, while methods such as photogrammetry, laser or structured-light scanner and the iPad-based scans capture only the external surfaces of the instruments, ($\mu$-)CT scans provide volumetric information. In this work, however, we restricted our analysis to the outer envelope of the instruments, following the delineation procedures for sound boards and backs described in \cite{beghin2023validation}. The second key difference lies in the format of the acquired data. Most techniques generate complete meshes, but the TLS provides only a point cloud. Hence, the sound board and back of the TLS acquisitions had to be delineated manually and not according to this mesh-based procedure. Furthermore, since point clouds can be derived from meshes, we chose to standardise all outputs in point cloud format in order to perform consistent comparisons across technologies.

\begin{figure}
    \centering
\includegraphics[width=\linewidth]{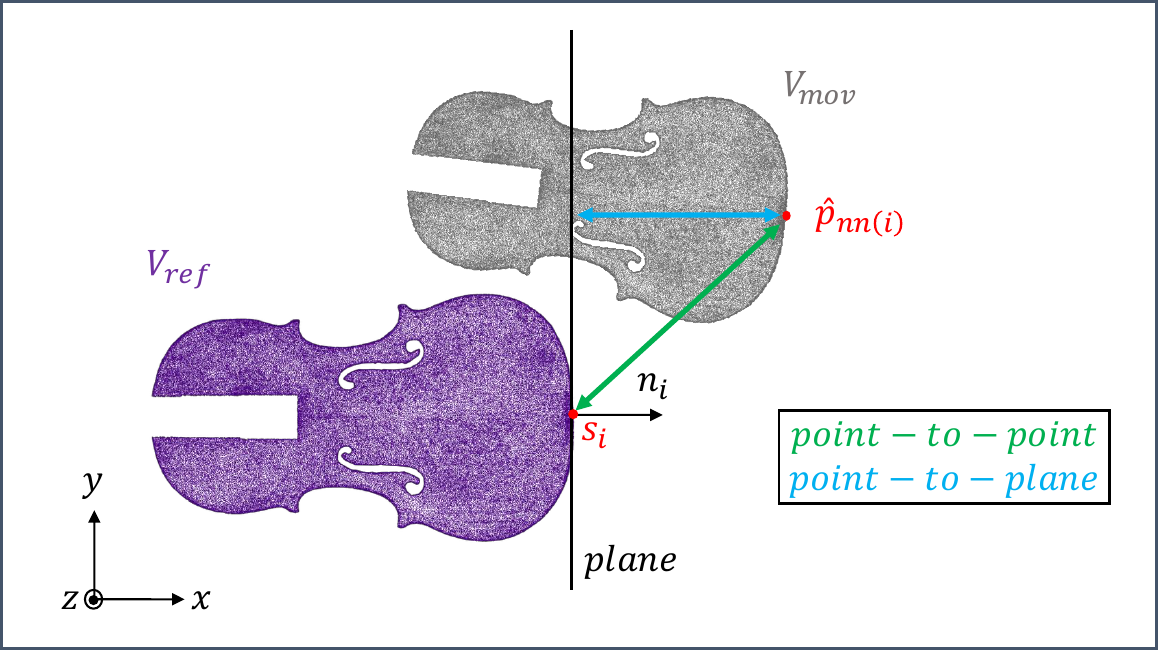}
    \caption{Difference between point-to-point and point-to-plane metrics}
    \label{fig:metrics}
\end{figure}

\noindent To evaluate the accuracy of the different digitisation methods, the point clouds from the same instrument were systematically compared against each other. Indeed, as no digital ground truth exists (i.e. no digital 3D model perfectly representing the instruments is available), we can only evaluate the discrepancy between the different clouds and attempt to quantify the error intrinsic to each technology. We perform a rigorous comparison between two techniques by evaluating the minimum average distance between two aligned point clouds of a same instrument acquired from these two techniques. Instead of relying on point-to-point distances\cite{besl1992method}, which are highly sensitive to local noise and variations in sampling density, we employed a point-to-plane metric as in \cite{chen1992object,beghin2023validation,terlindencomparison} for the registration (i.e. the alignment) of the two compared point clouds. This metric computes the distance from each point in a cloud to the local surface of the reference cloud, approximated by a plane defined by its nearest neighbours. This method provides a more robust estimation of deviations between models and mitigates the impact of noise. Although point clouds do not inherently provide surface normals, these can be estimated by applying principal component analysis (PCA) to the covariance matrix of the coordinates of neighboring points \cite{shakarji1998least,glira2015a}. The difference between point-to-point and point-to-plane metrics is illustrated in Figure \ref{fig:metrics}, adapted from \cite{beghin2023validation} (where mathematical details on both metrics are provided). Note that this distance between two clouds is not symmetric, as it depends on which point cloud is selected as the reference. This allows to double the number of comparisons. 

\noindent Since each acquisition technique produces already scaled models, omitting scaling in the point clouds registration is physically justified. Allowing scaling could artificially distort this property and bias the computed distance between two models based on different techniques. Therefore, we used a state of the art implementation of \texttt{SimpleICP} \cite{glira2015a}, which performs point-to-plane registration with internal point subsampling and without scaling. For comparable accuracy (from a thousandth to a hundredth of \SI{}{\mm}), this method is considerably faster than the point-to-plane approach developed in \cite{beghin2023validation}, reducing computation time from several minutes to just a few seconds for typical datasets. Its speed and stability make it particularly well suited for large-scale comparisons across acquisition techniques. 

\noindent Tables \ref{tab:acquisitions} and \ref{tab:compared technologies} summarise which techniques were employed to acquire each instrument, hence determining which techniques will be compared to each other. In all cases, both the sound board and the back were acquired, except for cello 1374, where only the sound board was captured using structured-light (both in blue-light and infrared modes). A total of 254 comparisons were performed. They are of the form of Equations \ref{eq:1}  or \ref{eq:2} (more on this in Sections \ref{section:NNLS} and \ref{section:NNLAD}). The detailed number of comparisons can be found per instrument in Table \ref{tab:acquisitions}, while Table \ref{tab:compared technologies} shows the number of times a technique is involved in the comparisons. The values in Table \ref{tab:compared technologies} sum up to 508, as each comparison is made between two technologies (each of the 254 equations contains two terms, which therefore sum up to 508). 

\begin{table}[H]
\centering
\small
\begin{tabular}{|l|c|c|c|c|c|}
\hline
\textbf{} & \textbf{1355} & \textbf{1374} & \textbf{1731} & \textbf{2020} & \textbf{2863} \\
\hline
\textbf{Photogrammetry} & \checkmark & \checkmark & \checkmark & \checkmark & \\
\hline
\textbf{Photog. PANORAMA} & \checkmark & & & & \checkmark \\
\hline
\textbf{Medical CT scan} & \checkmark & & & & \checkmark  \\
\hline
\textbf{Industrial $\mu$-CT scan} & & & \checkmark & & \\
\hline
\textbf{Structured-light BL} & & \checkmark & \checkmark & \checkmark & \\
\hline
\textbf{Structured-light IR} & \checkmark & \checkmark & & & \checkmark \\
\hline
\textbf{TLS} & \checkmark & & & & \checkmark \\
\hline
\textbf{LiDAR (Scaniverse)} & \checkmark & \checkmark & & & \checkmark \\
\hline
\textbf{LiDAR (Heges)} & \checkmark & \checkmark & & & \checkmark \\
\hline
\textbf{TrueDepth} & \checkmark & \checkmark & & & \checkmark \\
\hhline{|=|=|=|=|=|=|}
\textbf{Total sound board (= 136)} & 56 & 30 & 6 & 2 & 42 \\
\hline
\textbf{Total back (= 118)} & 56 & 12 & 6 & 2 & 42 \\
\hline
\textbf{Total comparisons (= 254)} & 112 & 42 & 12 & 4 & 84 \\
\hline
\end{tabular}
\caption{Summary of the digitised instruments. All acquisitions include the sound board and back except for instrument 1374 where only the sound board was acquired using structured-light. The last three lines indicate the number of systematic comparisons involving each plate for each instrument, summing up to a total of 254 comparisons.}
\label{tab:acquisitions}
\end{table}

\begin{table*}
\centering
\resizebox{\textwidth}{!}{
\begin{tabular}{|l|c|c|c|c|c|c|c|c|c|c|c|}
\hline
\textbf{} & \textbf{Photog.} & \textbf{Photog. PAN.} & \textbf{CT} & \textbf{$\mu$-CT} & \textbf{SL BL} & \textbf{SL IR} & \textbf{TLS} & \textbf{LiDAR Scan.} & \textbf{LiDAR Heges} & \textbf{TrueDepth} &\textbf{Tot.} \\
\hline
\textbf{Photogrammetry} & \cellcolor{lightgray} & \checkmark & \checkmark & \checkmark & \checkmark & \checkmark & \checkmark & \checkmark & \checkmark &\checkmark & 56 \\
\hline
\textbf{Photogrammetry PANORAMA} & \checkmark & \cellcolor{lightgray} & \checkmark & \cellcolor{lightgray} & \cellcolor{lightgray} & \checkmark & \checkmark &\checkmark & \checkmark & \checkmark & 52 \\
\hline
\textbf{Medical CT scan} & \checkmark & \checkmark & \cellcolor{lightgray} & \cellcolor{lightgray} & \cellcolor{lightgray} & \checkmark & \checkmark & \checkmark & \checkmark & \checkmark & 52 \\
\hline
\textbf{Industrial $\mu$-CT scan} & \checkmark & \cellcolor{lightgray} & \cellcolor{lightgray} & \cellcolor{lightgray} & \checkmark & \cellcolor{lightgray} & \cellcolor{lightgray} & \cellcolor{lightgray} & \cellcolor{lightgray} & \cellcolor{lightgray} & 8 \\
\hline
\textbf{Structured-light Blue-Light} & \checkmark & \cellcolor{lightgray} & \cellcolor{lightgray} & \checkmark & \cellcolor{lightgray} & \checkmark & \cellcolor{lightgray} & \cellcolor{lightgray} & \checkmark & \cellcolor{lightgray} & 22 \\
\hline
\textbf{Structured-light Infrared} & \checkmark & \checkmark & \checkmark & \cellcolor{lightgray} & \checkmark & \cellcolor{lightgray} & \checkmark & \checkmark & \checkmark & \checkmark & 62 \\
\hline
\textbf{TLS} & \checkmark & \checkmark & \checkmark & \cellcolor{lightgray} & \cellcolor{lightgray} & \checkmark & \cellcolor{lightgray} & \checkmark & \checkmark &\checkmark & 52 \\
\hline
\textbf{LiDAR Scaniverse} & \checkmark & \checkmark & \checkmark & \cellcolor{lightgray} & \cellcolor{lightgray} & \checkmark & \checkmark & \cellcolor{lightgray} & \checkmark &\checkmark & 68 \\
\hline
\textbf{LiDAR Heges} & \checkmark & \checkmark & \checkmark & \cellcolor{lightgray} & \checkmark & \checkmark & \checkmark & \checkmark & \cellcolor{lightgray} & \checkmark & 68 \\
\hline
\textbf{TrueDepth} & \checkmark & \checkmark & \checkmark & \cellcolor{lightgray} & \cellcolor{lightgray} & \checkmark & \checkmark & \checkmark & \checkmark & \cellcolor{lightgray} & 68 \\
\hline
\end{tabular}
} 
\caption{Summary of the compared technologies in this paper. Per line, the last value indicates how many times the technique on the left appears in the 254 comparisons.}
\label{tab:compared technologies}
\end{table*}

\noindent In this work we assume that the overall distance between two digital representations reflects the combined contributions of both acquisition techniques. We investigate this hypothesis through two complementary approaches. Both are based on the average squared point-to-plane distance $D^2_{plane}$ calculated by \texttt{SimpleICP}. More precisely, \texttt{SimpleICP} aligns two point clouds by minimising the sum of squared point-to-plane distances between the points of the reference cloud and their corresponding points in the moving cloud, and $D^2_{plane}$ is the resulting average squared distance. We then use the square root of this quantity, expressed in millimetres, for better readability. The 254 computed point-to-plane distances $\sqrt{D^2_{plane}}$ can be found in Tables \ref{tab:pairwise_distances_1} and \ref{tab:pairwise_distances_2} in Appendix \ref{section:appendix}.

\subsection{Non-Negative Least Squares (NNLS)}
\label{section:NNLS}

\noindent In a first formulation, we consider a model where the measured point-to-plane distance $\sqrt{D^2_{plane}}$ between two point clouds is expressed as the sum of the intrinsic error associated with each of the two acquisition techniques being compared. For each of the 254 pairwise comparisons, the model reads
\begin{equation}
\label{eq:1}
\sqrt{D^2_{plane}(\text{tech}_i,\text{tech}_j)} = e_{\text{tech}_i} + e_{\text{tech}_j} + \varepsilon_{ij},
\end{equation}
where $e_{\text{tech}_i}$ and $e_{\text{tech}_j}$ denote the intrinsic errors of techniques $i$ and $j$, respectively, and $\varepsilon_{ij}$ is a residual term accounting for measurement noise and
model mismatch. 

\noindent
\textbf{Per plate NNLS estimation.} We first perform a separate estimation for each plate. Let $\mathbf{e}^{(p)} = (e^{(p)}_1, \dots, e^{(p)}_{N_p})^\top$ be the vector of intrinsic errors for the $N_p$ compared techniques for a given plate $p$ (with $N_p$ ranging from 3 to 8), and $\mathbf{d}^{(p)}$ the vector collecting all corresponding measured distances $\sqrt{D^2_{plane}}$, whose length is therefore the number of comparisons $C_p = N_p (N_p-1)$ for plate $p$. The estimation problem can be written in matrix form as
\begin{equation*}
\mathbf{A}^{(p)}\mathbf{e}^{(p)} + \boldsymbol{\varepsilon} = \mathbf{d}^{(p)},
\end{equation*}
where each row of matrix $\mathbf{A^{(p)}}$ contains exactly two non-zero coefficients equal to one, corresponding to the two techniques involved in the measurement. We perform a weighted non-negative least-squares (NNLS) estimation of each vector $\mathbf{e}^{(p)}$, where  intrinsic errors are constrained to be non-negative, as they represent magnitude quantities. 
Our estimation procedure will minimise the avertage (or, equivalently, the sum) of squared residuals, as follows 
\begin{equation}
\hat{\mathbf{e}}^{(p)} = \arg\min_{\mathbf{e}^{(p)} \ge 0} \frac{1}{C_p} \sum_{k=1}^{C_p}
\left( \varepsilon_k \right)^2 = \arg\min_{\mathbf{e}^{(p)} \ge 0} \frac{1}{C_p} \sum_{k=1}^{C_p}
\left( \mathbf{A}^{(p)}_k \mathbf{e}^{(p)} - d^{(p)}_k \right)^2.
\tag{\(NNLS\)}
\label{eq:NNLS}
\end{equation}
Results are reported per plate in Table \ref{tab:intrinsic_errors_nnls}, where instruments are ordered by size (1355 being the smallest and 1374 the largest). As instrument 2020 only involves two pairwise comparisons per plate (photogrammetry versus structured-light in blue-light mode), it is not possible to solve a least-squares problem to estimate the intrinsic error for those two acquisition techniques. In this case, we can only report the measured distances (see Table \ref{tab:pairwise_distances_1} in Appendix \ref{section:appendix}), and cannot attribute a specific portion of the discrepancy to either technology. 

\noindent \textbf{Global WNNLS estimation.}  The last column of Table \ref{tab:intrinsic_errors_nnls} contains results of the NNLS applied to the 254 available comparisons at once, i.e. using a single matrix $\mathbf{A} \in \mathbb{R}^{254 \times 10}$ and measurement vector $\mathbf{d} \in \mathbb{R}^{254}$ that covers all plates. Due to the uneven representation of techniques in the dataset, a standard least-squares formulation would overweight techniques appearing in a large number of pairwise comparisons (e.g. numerous iPad acquisitions vs. few $\mu$-CT acquisitions). To mitigate this effect, we adopt a weighted non-negative least-squares approach (WNNLS). Let $c_i$ denote the number of times technique $i$ appears in the system of 254 equations. For a measurement involving techniques $i$ and $j$, the associated weight is defined as
\begin{equation*}
w_k = \frac{1}{2}\left(\frac{1}{c_i} + \frac{1}{c_j}\right),
\label{eq:W}
\end{equation*}
which increases the influence of equations involving sparsely represented techniques while preventing heavily sampled techniques from dominating the estimation. The resulting minimisation problem now considers the full set of equations $\mathbf{A}\mathbf{e} + \boldsymbol{\varepsilon} = \mathbf{d}$ and can be written
\begin{equation}
\hat{\mathbf{e}} = \arg\min_{\mathbf{e} \ge 0} \frac{1}{C_p} \sum_{k=1}^{254}
w_k\left( \mathbf{A}_k \mathbf{e} - d_k \right)^2,
\tag{\(WNNLS\)}
\label{eq:WNNLS}
\end{equation}
where $w_k$ is the weight assigned to the $k$-th equation.

\noindent The techniques listed in each row of Table \ref{tab:intrinsic_errors_nnls} are ranked in ascending order of intrinsic error according to the WNNLS criterion applied to all 254 comparisons. The smallest error is attributed to $\mu$-CT and the largest to the iPad LiDAR acquisition using the Heges application. Results for which the individual ranking per plate does not follow the overall ranking are highlighted in bold. An alternative, more visual representation of the results of Table \ref{tab:intrinsic_errors_nnls} is shown in Figure \ref{fig:Error_NNLS}, where segments represent the range of the different errors computed, a triangle the intrinsic error of the instruments acquisitions when calculated by plate, and a cross the value calculated on the basis of all 254 comparisons using weighted non-negative least squares.

\noindent Finally, the last line at the bottom of Table \ref{tab:intrinsic_errors_nnls} shows the Root Mean Square (RMS) of the residuals $\boldsymbol{\varepsilon}$, i.e. the square root (in \SI{}{\mm}) of the optimal values of the objective functions of (NNLS) and (WNNLS). It indicates the average residual error in each individual equation used to evaluate the intrinsic error of each technology. The smaller the RMS, the more accurate the model. On the other hand, a greater RMS might imply an important  measurement noise, an ill-fitting model or the presence of outliers. 


\subsection{Non-Negative Least Absolute Deviation (NNLAD)}
\label{section:NNLAD}

\noindent Using a least-squares objective to fit a model is a standard approach for estimation, which was used in the previous section to fit an additive model for the point-to-plane distances. However, it can be argued that, in a Euclidean space, a more natural model should be additive in squared distances, as the Pythagorean theorem suggests. Hence we  present here the following alternative model based on squared distances:
\begin{equation}
\label{eq:2}
D^2_{plane}(\text{tech}_i,\text{tech}_j) = e_{\text{tech}_i}^2 + e_{\text{tech}_j}^2 + \tilde{\varepsilon}_{ij},
\end{equation}
where the squared intrinsic errors are assumed to contribute additively to the observed squared distance. In this case, the matrix formulation is 
\begin{equation*}
\mathbf{A}\mathbf{e}^2 + \boldsymbol{\tilde{\varepsilon}} = \mathbf{d}^2,
\end{equation*}
(where the square on vectors is applied componentwise)
and the estimation problem is formulated using a least average tabsolute deviation criterion $ \frac{1}{C_p} \sum_{k=1}^{C_p}  \left| \tilde{\varepsilon}_k \right| = \frac{1}{C_p} \| \boldsymbol{\tilde{\varepsilon}} \|_1$ instead of least-squares
\begin{equation}
\hat{\mathbf{e}}^{(p)} =  \sqrt{\arg\min_{\mathbf{e}^{(p)}{}^2 \ge 0} \frac{1}{C_p} \sum_{k=1}^{C_p} \left| \mathbf{A}^{(p)}_k\mathbf{e}^{(p)}{}^2 - d^{(p)}_k{}^2\right|},
\tag{\(NNLAD\)}
\label{eq:NNLAD}
\end{equation}
which aims at reducing the influence of outliers and provides a more robust estimate when deviations from the model assumptions occur. Results are reported per plate in Table \ref{tab:intrinsic_errors_nnlad}. The last column contains results of the NNLAD applied to the 254 comparisons at once, which have also been weighted due to the uneven representation of acquisition techniques, corresponding to: 
\begin{equation}
\hat{\mathbf{e}} = \sqrt{\arg\min_{\mathbf{e}{}^2 \ge 0} \frac{1}{254} \sum_{k=1}^{254} w_k \left| \mathbf{A}_k\mathbf{e}^2 - d^2_k\right|},
\tag{\(WNNLAD\)}
\label{eq:WNNLAD}
\end{equation}
with the weights $w_k$ defined as in the WNNLS approach\footnote{Although both weighted formulations involve the same coefficients $w_k$, their numerical implementation differs due to the nature of the objective functions. In the WNNLS formulation, $\sum_{k=1}^{254} w_k \left( \mathbf{A}_k \mathbf{e} - d_k \right)^2$, the residuals enter quadratically. To obtain an equivalent standard least-squares problem, each equation is therefore scaled by $\sqrt{w_k}$, so that the solver minimises $\|\sqrt{\mathbf W}(\mathbf A\mathbf e-\mathbf d)\|_2^2$. In practice, this amounts to premultiplying both $\mathbf A$ and
$\mathbf d$ by $\sqrt{w_k}$. In contrast, the WNNLAD formulation minimises a weighted $\ell_1$-norm, $\sum_{k=1}^{254} w_k \left| \mathbf{A}_k\mathbf{e}^2 - d_k^2 \right|$. Since the objective is linear in the residual magnitude, the correct weighting is obtained by scaling each equation by $w_k$, without taking the square root. The use of $\sqrt{w_k}$ for NNLS and $w_k$ for NNLAD thus follows naturally from the algebraic structure of the respective optimisation problems and ensures consistency between the mathematical formulations and
their numerical implementation.}. A visual representation of the results can be found in Figure \ref{fig:Error_LAD}. 

\noindent The first of the last two lines of Table \ref{tab:intrinsic_errors_nnlad} shows the average $\ell_1$-norm of the residuals $\boldsymbol{\tilde{\varepsilon}}$. This value is harder to interpret than the RMS of the residuals $\boldsymbol{\varepsilon}$ in the case of the NNLS metric because it is expressed in \SI{}{\mm^2}, hence we also indicate its square root (in \SI{}{\mm}) on the last line, for better readability.

\section{Results}

\begin{table*}
\centering
\begin{tabular}{lccccccccc}
\toprule
\textbf{Technique \quad \textbackslash \quad Plate} &
\textbf{1355 sb} &
\textbf{1355 back} &
\textbf{1731 sb} &
\textbf{1731 back} &
\textbf{2863 sb} &
\textbf{2863 back} &
\textbf{1374 sb} &
\textbf{1374 back} &
\textbf{All} \\
\midrule
Industrial $\mu$-CT scan
& - & -
& \textbf{0.080} & \textbf{0.059}
& - & -
& - & -
& 0.000 \\

Structured-light (BL)
& - & -
& \textbf{0.000} & \textbf{0.050}
& - & -
& 0.021 & -
& 0.058 \\

Photog. PANORAMA
& \textbf{0.109} & \textbf{0.066}
& - & -
& 0.281 & 0.356
& - & -
& 0.166 \\

Structured-light (IR)
& \textbf{0.054} & \textbf{0.063}
& - & -
& 0.397 & \textbf{0.546}
& \textbf{0.337} & -
& 0.275 \\

Photogrammetry
& \textbf{0.355} & \textbf{0.231}
& 0.134 & 0.060
& - & -
& \textbf{0.758} & \textbf{0.820}
& 0.294 \\

TrueDepth
& \textbf{0.124} & \textbf{0.136}
& - & -
& \textbf{0.574} & \textbf{0.596}
& \textbf{0.046} & \textbf{0.367}
& 0.342 \\

TLS
& \textbf{0.557} & 0.406
& - & -
& \textbf{0.419} & \textbf{0.320}
& - & -
& 0.409 \\

Medical CT scan
& \textbf{0.388} & \textbf{0.482}
& - & -
& 0.697 & 0.900
& - & -
& 0.578 \\

LiDAR (Scaniverse)
& 0.803 & \textbf{0.456}
& - & -
& 1.251 & 1.202
& 1.063 & 1.167
& 0.987 \\

LiDAR (Heges)
& 1.543 & 1.393
& - & -
& 1.454 & 1.701
& 2.479 & 1.858
& 1.724 \\
\bottomrule

$\sqrt{\frac{1}{C_p}\sum_{k=1}^{C_p}
\left( \varepsilon_k \right)^2} \hspace{0.2cm} [mm]$
& 0.348 & 0.238
& 0.080 & 0.011
& 0.262 & 0.402
& 0.825 & 0.679
& 0.069 \\

\bottomrule
\end{tabular}
\caption{Estimated intrinsic errors (in mm) for each acquisition technique (when available) on each plate using non-negative Least-Squares methods (NNLS). The notation `sb' refers to the sound board, while `back' explicitly indicates the back. Instruments and acquisition techniques are ordered by increasing instrument size from left to right, and by increasing intrinsic error from top to bottom. Instances in which this ranking is not preserved and where techniques should be interchanged are highlighted in bold. The final column reports the intrinsic error estimated for each acquisition technique when a Weighted Non-Negative Least-Squares (WNNLS) model is applied jointly to the 254 pairwise comparisons. The last line indicates the value of the root mean square of the residuals $\boldsymbol{\varepsilon}$ (in mm).}
\label{tab:intrinsic_errors_nnls}
\end{table*}

\begin{table*}
\centering
\begin{tabular}{lccccccccc}
\toprule
\textbf{Technique\quad \textbackslash \quad Plate} &
\textbf{1355 sb} &
\textbf{1355 back} &
\textbf{1731 sb} &
\textbf{1731 back} &
\textbf{2863 sb} &
\textbf{2863 back} &
\textbf{1374 sb} &
\textbf{1374 back} &
\textbf{All} \\
\midrule
Industrial $\mu$-CT scan
& -     & -
& 0.046 & \textbf{0.088}
& -     & -
& -     & -
& 0.000 \\

Structured-light (BL)
& - & -
& 0.053 & \textbf{0.052}
& -     & -
& 0.331 & -
& \textbf{0.128} \\

Photog. PANORAMA
& \textbf{0.044} & 0.000
& -     & -
& 0.218 & \textbf{0.525}
& -     & -
& \textbf{0.000} \\

Structured-light (IR)
& \textbf{0.000} & 0.000
& -     & -
& 0.504 & \textbf{0.605}
& \textbf{0.395} & -
& \textbf{0.263} \\

Photogrammetry
& 0.193 & 0.148
& 0.071 & \textbf{0.077}
& -     & -
& \textbf{0.549} & 0.644
& \textbf{0.121} \\

TrueDepth
& 0.211 & 0.223
& -     & -
& \textbf{0.848} & \textbf{0.727}
& \textbf{0.358} & 0.657
& 0.464 \\

TLS
& \textbf{0.822} & 0.504
& -     & -
& \textbf{0.593} & \textbf{0.398}
& -     & -
& 0.687 \\

Medical CT scan
& \textbf{0.549} & \textbf{0.741}
& -     & -
& 1.045 & 1.263
& -     & -
& 0.771 \\

LiDAR (Scaniverse)
& 1.029 & \textbf{0.587}
& -     & -
& 1.530 & 1.477
& 1.167 & 1.363
& 1.206 \\

LiDAR (Heges)
& 1.537 & 1.434
& -     & -
& 1.721 & 1.960
& 2.016 & 1.903
& 1.817 \\

\bottomrule

$\frac{1}{C_p}\|\boldsymbol{\tilde{\varepsilon}}\|_1 \hspace{1.2cm} [mm^2]$
& 0.474 & 0.317
& 0.023 & 0.002
& 0.614 & 1.148
& 2.566 & 2.523
& 0.017 \\

$\sqrt{\frac{1}{C_p}\|\boldsymbol{\tilde{\varepsilon}}\|_1} \hspace{0.9cm} [mm]$
& 0.689 & 0.563
& 0.152 & 0.044
& 0.784 & 1.071
& 1.602 & 1.588 
& 0.132  \\

\bottomrule
\end{tabular}
\caption{Estimated intrinsic errors (in mm) for each acquisition technique (when available) on each plate using Non-Negative Least Absolute Deviation (NNLAD) with squared-distance formulation. The notation `sb' refers to the sound board, while `back' explicitly indicates the back. Instruments and acquisition techniques are ordered by increasing instrument size from left to right, and by increasing intrinsic error from top to bottom. Instances in which this ranking is not preserved and where techniques should be interchanged are highlighted in bold. The final column reports the intrinsic error estimated for each acquisition technique when a Weighted Non-Negative Least Absolute Deviation (WNNLAD) model is applied jointly to the 254 pairwise comparisons. The first of the last two lines indicates the average of the absolue values of the residuals $\boldsymbol{\tilde{\varepsilon}}$ (in \SI{}{\mm^2}) while the second one indicates its square root, for better interpretability (in \SI{}{\mm}).}
\label{tab:intrinsic_errors_nnlad}
\end{table*}

\begin{figure*}[p]
    \centering
    \begin{subfigure}[t]{\textwidth}
        \centering
        \includegraphics[width=\textwidth]{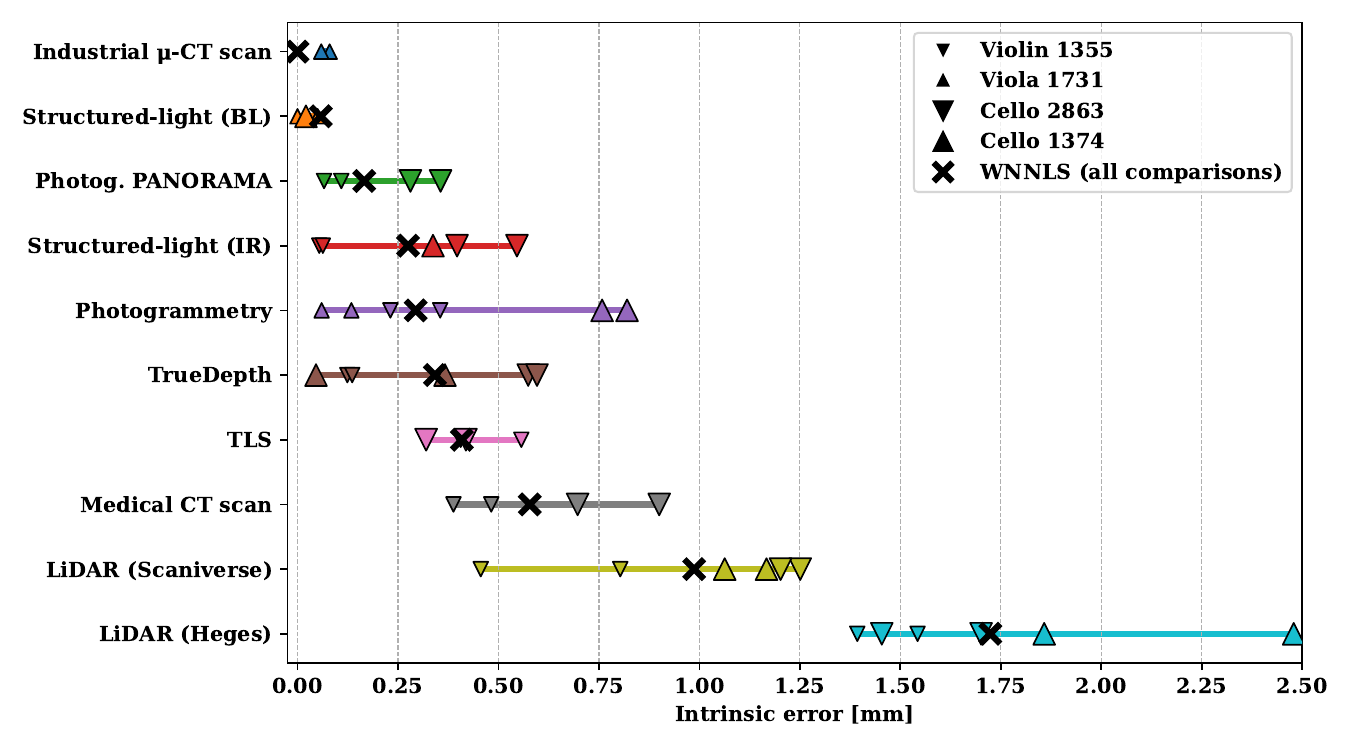}
        \caption{NNLS metric}
        \label{fig:Error_NNLS}
    \end{subfigure}

    \vspace{0.5em} 

    \begin{subfigure}[t]{\textwidth}
        \centering
        \includegraphics[width=\textwidth]{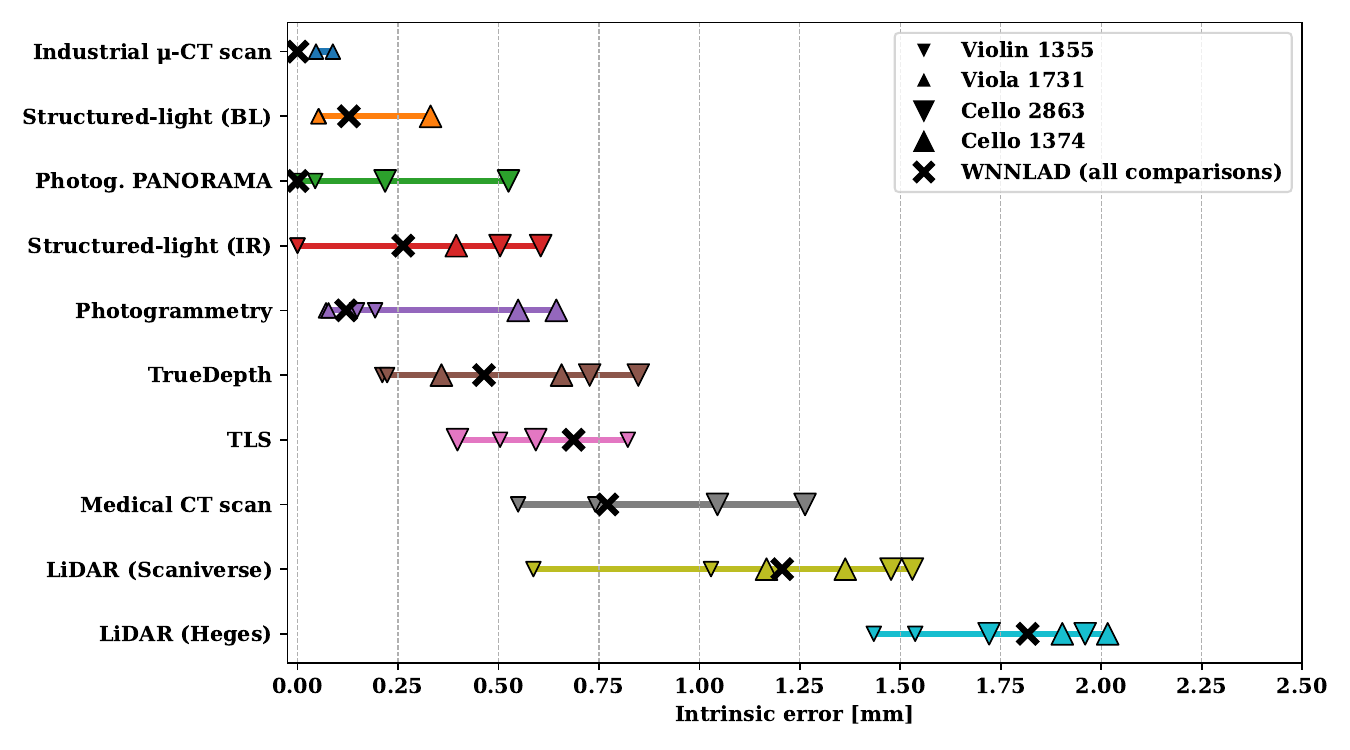}
        \caption{NNLAD metric}
        \label{fig:Error_LAD}
    \end{subfigure}
    \caption{Intrinsic errors per acquisition technique. Segments represent the range of the different errors computed, a triangle the intrinsic error of the instruments when calculated by plate, and a cross the value calculated on the basis of all 254 comparisons using weighted approach.}
    \label{fig:Errors_composite}
\end{figure*}

\noindent Results of the comparisons are represented in Tables \ref{tab:intrinsic_errors_nnls} and \ref{tab:intrinsic_errors_nnlad}, and more visually in Figure \ref{fig:Errors_composite}. A first observation emerging from these is the almost systematic difference in intrinsic error between smaller instruments (violin 1355 and viola 1731) and larger ones (cellos 2863 and 1374). This effect is observed for both metrics considered in this study, namely least squares and least absolue deviation. Technologies used with larger instruments tend to exhibit higher intrinsic errors than with smaller ones. Nevertheless, since all acquisition techniques were applied to both small and large instruments (except for $\mu$-CT scanning), the overall ranking of techniques remains coherent across instrument sizes. In other words, had the analysis been restricted exclusively to small or to large instruments, the general trends observed in the rankings would not have changed (one might have expected a slightly larger error for the $\mu$-CT scanning if both sizes were compared). In the global rankings produced by both metrics, $\mu$-CT scanning yields an intrinsic error of \SI{0.000}{\mm}. We caution the reader against a literal interpretation of this result: although $\mu$-CT is indeed known to be highly accurate according to the state of the art, it is also the least represented technique in our dataset, appearing in only 4 equations out of the 254 pairwise comparisons. Despite the weighting introduced in the WNNLS and WNNLAD formulations to mitigate such imbalances, both metrics converge towards a null intrinsic error for $\mu$-CT, which should be interpreted as a limitation of the available data rather than as an indication of perfect accuracy in practice. Tables \ref{tab:intrinsic_errors_nnls} and \ref{tab:intrinsic_errors_nnlad} suggest that the intrinsic error of $\mu$-CT would typically lie around a few hundredth of \SI{}{\mm}. Also, the value of the RMS value of the residuals $\boldsymbol{\varepsilon}$ (last line, Table \ref{tab:intrinsic_errors_nnls}) or the root average $\ell_1$-norm of the residuals $\boldsymbol{\tilde{\varepsilon}}$ (last line, Table \ref{tab:intrinsic_errors_nnlad}) provides a good indicator of the fit when compared to the values of the computed intrinsic errors of the two tables. In the case of the back of instrument 1731, we see that, for both metrics, the average residual is smaller than the other three intrinsic errors computed. We might then expect a good prediction. On the other hand, the average residual for the sound board or the back of instrument 1374 is almost always higher than the computed intrinsic errors, which might imply more measurement noise or outlier values. 

\noindent Beyond this size-related effect, the rankings produced by the two metrics are nearly identical and remarkably consistent with expectations derived from the state of the art and from the accuracy specifications reported by manufacturers. Most of the techniques show a sub-millimetre accuracy, except for the two LiDAR iPad-based methods. When using both metrics, $\mu$-CT scanning is identified as the technique with the smallest intrinsic error, while the largest errors are consistently associated, in increasing order, with TrueDepth, TLS, CT, and the two LiDAR iPad-based acquisitions. The relative positioning of structured-light scanning and photogrammetry is more nuanced and warrants closer examination. In both rankings, the photogrammetry with cross-polarisation performed by the PANORAMA platform yields lower intrinsic errors than the photogrammetric acquisitions without polarisation, and blue-light structured-light scanning consistently outperforms its infrared counterpart. However, the relative ordering between structured-light scanning and photogrammetry is not strictly identical across the two metrics. In one case, an alternation is observed between structured-light (blue light), photogrammetry by the PANORAMA platform, structured-light (infrared), and the other photogrammetric acquisitions, whereas in the other case, both photogrammetric approaches are estimated to be more accurate than structured-light scanning. In addition, Table \ref{tab:intrinsic_errors_nnlad} shows that the calculation of intrinsic error can still sometimes be sensitive to the metric used. Indeed, the accuracy of Structured-light (BL) and TrueDepth for instrument 1374 sound board shows an error of hundredths of a millimetre with NNLS, but of tenths of millimetres with NNLAD. This discrepancy highlights the sensitivity of overall rankings to the chosen error metric, while reinforcing the robustness of the broader trends observed across acquisition technologies. 

\noindent We observe that CT scanning does not perform as well as we assumed in our initial perception of its accuracy. In previous studies \cite{beghindigital,beghin2023validation}, we hypothesised that CT acquisitions would necessarily be more accurate than photogrammetric reconstructions, but the present results contradict this assumption. Similar conclusions have been reported in a related comparative study~\cite{mathys2013comparing}, where authors state that they did not use the CT model as a reference for their comparisons because this 3D model was the result of 2D image segmentation using a threshold value. This remark  also applies to our own workflow, in which two-dimensional CT slices were converted into point clouds using the 3D Slicer software. Somewhat unexpectedly, TLS, although primarily designed for the documentation of large-scale archaeological sites or objects of several tens of metres, exhibits a lower intrinsic error than CT, which is typically intended for medical acquisitions at the decimetre to metre scale. We attribute this relatively low TLS error mainly to the extensive cleaning and post-processing steps applied to the TLS models, as described in Section \ref{section:3D acquisition campaigns}. In particular, numerous spurious elements and a pronounced `quadruple' thickness artefact were removed, resulting in a substantially purified model. 

\noindent Then, although the state of the art generally suggests that structured-light scanning achieves higher accuracy than photogrammetry, we see that the photogrammetric acquisitions performed by the PANORAMA platform outperformed the infrared structured-light acquisitions, and even the blue-light mode when evaluated with the NNLAD metric. While this result may appear counterintuitive, it can be explained by several factors. Infrared structured-light scanning (already less accurate than blue-light projection) is more sensitive to complex geometries and surface reflections, both of which degraded the quality of the generated models in this case. 

\noindent It appears that the TrueDepth technology exhibits a relatively small intrinsic error. Given that this technology relies on a widely accessible consumer hardware, namely an iPad, substantially poorer performance had initially been expected, especially when compared to industrial acquisition methods. In retrospect, however, this result is perhaps less surprising than it first appears: TrueDepth was originally engineered for subtle facial recognition, a task that demands geometric sensitivity at short range, which may explain its ability to capture surface details with relatively high fidelity. Its performance indeed proves to be comparable to certain structured-light acquisitions performed in infrared mode (and was even better in some cases). A major drawback of this technique should nevertheless be noted: TrueDepth relies on the front-facing camera of the iPad, which makes it difficult for the operator to accurately assess the scanned area during acquisition, as the screen is located on the same side of the device. 

\noindent Finally, as expected, the results obtained using the iPad LiDAR technologies are less accurate than those acquired with other techniques. Nevertheless, we would like to point out that their intrinsic error lies in the order of millimetres, and sometimes even sub-millimetres for small instruments.  

\noindent Despite the care taken during acquisition and processing, these results remain subject to several limitations. It was not possible to digitise all instruments using all acquisition techniques, thereby restricting the scope of direct comparisons (as observed for the $\mu$-CT scanner). In addition, several regions of the models required manual cleanup deletions, introducing a degree of human intervention into the estimation of intrinsic errors. These include the f-holes and fingerboard regions, the quadruple layer artefact in TLS data and the double edge on structured-light (infrared mode) models, and internal surface layers present in CT and $\mu$-CT reconstructions. It should also be emphasised that the trends discussed above are valid for the specific set of instruments considered in this study, namely violins, violas, and cellos. These objects, with characteristic dimensions on the order of one metre, have highly specific geometries, and the analysed plates (sound board and back) can be considered quasi-planar surfaces. Consequently, the conclusions drawn here may not be directly applicable to objects of substantially different sizes or geometric characteristics. Indeed, it is possible that the actual accuracy of certain acquisition techniques is underestimated or, conversely, appears improved when evaluated solely on these restricted regions. Typically, each plate was acquired three times using the iPad technologies (LiDAR on Scaniverse and Heges, as well as TrueDepth), but only the best acquisition was kept. Finally, we remain cautious in the interpretation of these numerical values. Indeed, we saw that due to the minimisation of residuals in the WNNLS and WNNLAD approaches, technologies with small errors such as $\mu$-CT scanning are considered perfect. It also appears that larger instruments tend to induce a greater intrinsic error to their acquisition technologies than smaller ones. Furthermore, violin plates share highly similar geometric characteristics, and we previously showed in \cite{beghin2023validation} that even comparisons between different instruments can yield relatively small discrepancies, only slightly exceeding \SI{1}{\mm}.

\section{Discussion}

\noindent The results presented in this study highlight the fact that no single acquisition technique can be considered universally optimal. In fact, the suitability of a given method strongly depends on the intended application, the required level of accuracy, the importance of surface appearance, and practical constraints such as cost, acquisition time, and accessibility. The observed ranking on the qualitative behaviour of the evaluated techniques in Tables \ref{tab:intrinsic_errors_nnls} and \ref{tab:intrinsic_errors_nnlad} is largely consistent with previous comparative studies and with the performance claims reported in the literature. Table \ref{tab:techniques_summary} at the end of the discussion summarises the main findings discussed hereafter. 

\noindent Photogrammetry emerges as a particularly attractive compromise between accuracy, cost, and accessibility, especially in museum and heritage contexts. It is well suited for the digitisation of medium- to large-sized cultural heritage objects, for which both geometry and visual appearance are of interest. A major advantage of photogrammetry is its ability to faithfully capture surface texture, which is essential for visual inspection, documentation and public dissemination, as emphasised in previous museum-oriented studies. Furthermore, the relatively low cost of the required hardware makes photogrammetry particularly appealing. 

\noindent However, photogrammetry also presents several limitations. First, it does not intrinsically provide models at an absolute scale and therefore requires the use of scale bars, markers, or reference measurements, introducing an additional source of potential human error. In contrast, techniques such as structured-light scanning, CT, and $\mu$-CT inherently produce metrically scaled models. Second, photogrammetry is highly sensitive to acquisition conditions, including lighting setup, surface reflectivity, and camera placement. Varnished or reflective materials, such as those commonly encountered on musical instruments, can significantly degrade reconstruction quality if diffuse lighting is not carefully controlled. Finally, the absence of real-time feedback during acquisition increases the risk of incomplete or suboptimal coverage, making the method particularly dependent on operator expertise. While photogrammetry is generally less expensive than structured-light systems, it may require substantial post-processing effort, especially when acquisition conditions are not optimal. 

\noindent The $\mu$-CT technique offers unmatched accuracy and remains the reference technique when the study of internal structures is required. Its ability to capture volumetric information makes it indispensable for applications such as material analysis, internal geometry inspection, and acoustic modelling that relies on thickness variations and internal features. Nevertheless, $\mu$-CT is resource-intensive, requiring specialised infrastructure, long acquisition times, and significant data processing. These constraints, combined with limited accessibility, restrict its use to specific research-driven applications rather than routine documentation or large-scale digitisation campaigns. 

\noindent Structured-light scanning represents a well-balanced alternative, combining relatively high accuracy with direct metric output and a more controlled acquisition process. Blue-light structured-light systems, in particular, achieved very low intrinsic errors in our study, confirming their suitability for high-precision surface digitisation. Compared to photogrammetry, structured-light scanning offers the advantage of immediate feedback and reduced reliance on external scaling. However, performance can degrade when dealing with complex geometries, reflective surfaces, or large objects acquired in a continuous scanning mode, where tracking losses may occur. Infrared structured-light scanning was observed to be less robust in such conditions, especially in the presence of strong reflections and challenging incidence angles. 

\noindent Mobile LiDAR solutions, such as those based on consumer-grade devices (e.g., iPad LiDAR using Scaniverse or Heges applications), stand out primarily for their speed and ease of use and relatively low cost. These techniques enable rapid acquisition with minimal setup, making them particularly suitable for rapid surveys, fieldwork, and educational or outreach applications. While their geometric accuracy remains significantly lower than that of photogrammetry, structured-light scanning, or CT-based techniques, their performance is nonetheless sufficient for low-fidelity models and contextual documentation. Among the tested applications, Heges appeared promising with the TrueDepth technology, although its reliance on the front-facing camera limits real-time visual control during acquisition, which can complicate the scanning process. 

\noindent Beyond purely geometric considerations, the relative importance of texture versus shape plays a decisive role in selecting an acquisition technique. Several studies have emphasised that when surface appearance and visual realism are prioritised, photogrammetry remains the method of choice, whereas laser-based techniques may be preferred when geometric fidelity alone is required. This distinction is particularly relevant in museum contexts, where textured 3D models support conservation workflows, visual inspection, and public dissemination, including augmented reality and interactive exhibits. Moreover, maintaining accurate digital surrogates of cultural artefacts provides a valuable reference for conservation purposes in the event of damage or degradation. 

\noindent In the specific context of this study, CT and TLS can be regarded as comparatively less competitive acquisition techniques. While CT scanning is often perceived as a high-precision reference method, it does not provide surface texture information and is particularly demanding in terms of implementation. As a result, the final surface models derived from CT did not achieve the level of geometric accuracy obtained with several surface-based acquisition techniques evaluated in this work, despite the considerable resources involved. Similarly, TLS proved poorly adapted to the digitisation of musical instruments in this experimental setup. The raw output consisted of a point cloud of very limited quality, likely reflecting the atypical use of TLS for objects of small size and complex geometry. In contrast to many surface acquisition techniques, TLS did not directly produce a usable mesh (only a point cloud), and automatic meshing approaches commonly available in dedicated software failed due to the presence of the `quadruple' thickness artefact. Addressing this issue required extensive manual intervention to clean and filter the point cloud in order to retain a single surface layer. This substantial human effort was necessary to achieve acceptable results, yet it did not ultimately lead to a clear improvement over the other acquisition techniques in terms of accuracy. In this respect, the performance of TLS in the present study appears largely dependent on post-processing rather than on the intrinsic suitability of the technique itself, thereby limiting its practical relevance for the digitisation of musical instruments.

\noindent Overall, these results reinforce the idea that the choice of an acquisition technique should be guided by clearly defined objectives rather than by an assumed hierarchy of accuracy. Different research contexts naturally favour different technologies. For applications such as acoustic modelling or internal structure studies, where access to volumetric information or precise thickness measurements is required, $\mu$-CT remains the most appropriate solution despite its cost and complexity. Conversely, for archival documentation and museum-oriented digitisation, photogrammetry and structured-light scanning emerge as particularly well-suited techniques, offering a favourable compromise between geometric accuracy, acquisition effort, and visual completeness. Photogrammetry is especially attractive when surface texture and visual realism are important, for instance in documentation, conservation records, or public dissemination, whereas structured-light scanning provides metrically scaled models with reduced sensitivity to operator-dependent scaling errors. For education, outreach, rapid surveys, or fieldwork, mobile LiDAR solutions such as iPad-based systems offer unmatched speed and accessibility, albeit at the cost of reduced geometric fidelity, making them better suited for low-fidelity or contextual models rather than precise measurements. Taken together, these observations underline that no single technique is universally optimal: instead, each method occupies a well-defined niche, and selecting the “right tool for the right purpose” is essential to balance accuracy, practicality, and intended use.

\begin{table*}[htbp]
\centering
\setlength{\tabcolsep}{3pt}
\renewcommand{\arraystretch}{1.20}
\scriptsize

\begin{tabular}{
p{2.5cm}
p{2.8cm}
p{2.5cm}
p{1.4cm}
p{2.6cm}
p{5.1cm}
}
\toprule

\textbf{Technique} &
\textbf{Physical principle} &
\textbf{Intrinsic accuracy range [mm]} &
\textbf{Scaling required} &
\textbf{Cost / accessibility} &
\textbf{Recommended use} \\

\midrule

Industrial $\mu$-CT &
\begin{tabular}[t]{@{}l@{}}X-rays\\Volumetric imaging\end{tabular} &
$\sim$0--0.1$^{a}$ &
No &
\begin{tabular}[t]{@{}l@{}}Very high cost\\Time-consuming\\Specialised infra-\\structure\\Staff demanding\\Demanding post-\\processing\end{tabular} &
\begin{tabular}[t]{@{}l@{}}Very high accuracy\\Internal structures and volumetric study\\Thickness measurements\\Acoustic modelling (in case of violins)\\Non-invasive measurements\end{tabular} \\
\addlinespace[6pt]

Structured-light (BL) &
\begin{tabular}[t]{@{}l@{}}Projected patterns\\Blue visible light\\Triangulation\end{tabular} &
0.06--0.13 &
No &
\begin{tabular}[t]{@{}l@{}}$\sim$\EUR{8--10k}\\Professional scanner\\Direct metric output\\Real-time feedback\end{tabular} &
\begin{tabular}[t]{@{}l@{}}High-precision surface digitisation\\Sensitive to reflective surfaces\\and complex geometries\\Museum documentation\\Non-invasive measurements \end{tabular} \\
\addlinespace[6pt]

Photogrammetry PANORAMA &
\begin{tabular}[t]{@{}l@{}}Visible light\\Cross-polarisation\end{tabular} &
0.00--0.17 &
Yes &
\begin{tabular}[t]{@{}l@{}}$\sim$\EUR{5k} (software and\\ hardware)\\Dedicated setup\\Demanding workflow\\and photo processing\end{tabular} &
\begin{tabular}[t]{@{}l@{}}Public dissemination\\High-fidelity surface documentation\\High-quality texture\\Suitable for varnished or\\reflective objects\\Museum documentation\\Non-invasive measurements \end{tabular} \\
\addlinespace[6pt]

Structured-light (IR) &
\begin{tabular}[t]{@{}l@{}}Projected patterns\\Infrared light\\Triangulation\end{tabular} &
0.26--0.28 &
No &
\begin{tabular}[t]{@{}l@{}}$\sim$\EUR{5k}\\Professional scanner\\Direct metric output\\Real-time feedback\end{tabular} &
\begin{tabular}[t]{@{}l@{}}Surface digitisation\\Useful when blue-light acquisition\\is impaired by surface properties\\Museum documentation\\Non-invasive measurements\end{tabular}\\
\addlinespace[6pt]

Photogrammetry &
\begin{tabular}[t]{@{}l@{}}Visible light\end{tabular} &
0.12--0.29 &
Yes &
\begin{tabular}[t]{@{}l@{}}$\sim$\EUR{5k} (software and\\ hardware)\\Camera and lighting\\ equipment\\Demanding workflow\\and photo processing\end{tabular} &
\begin{tabular}[t]{@{}l@{}}Public dissemination\\High-quality surface texture\\Sensitive to acquisition conditions\\Museum documentation\\Non-invasive measurements\end{tabular} \\
\addlinespace[6pt]

TrueDepth (Heges) &
\begin{tabular}[t]{@{}l@{}}Infrared struc-\\tured light\\IR camera + dot pro-\\jector\end{tabular} &
0.34--0.46 &
No &
\begin{tabular}[t]{@{}l@{}}$\sim$\EUR{1k}\\Consumer iPad Pro\\ Real-time feedback\\Fast and accessible\end{tabular}&
\begin{tabular}[t]{@{}l@{}}Rapid low-cost acquisition\\Contextual documentation\\Moderate geometric accuracy\\Education and outreach\\Front-facing sensor complicates\\real-time monitoring\end{tabular} \\
\addlinespace[6pt]

TLS &
\begin{tabular}[t]{@{}l@{}}Laser (near-infrared)\\Time-of-flight\end{tabular} &
0.41--0.69 &
No &
\begin{tabular}[t]{@{}l@{}}$\sim$\EUR{80k}\\Specialised equipment\\Substantial post-\\processing required\end{tabular} &
\begin{tabular}[t]{@{}l@{}}Large-scale surveying\\Architecture\\Archaeological sites\\Poorly adapted to musical instruments\end{tabular} \\
\addlinespace[6pt]

Medical CT scan &
\begin{tabular}[t]{@{}l@{}}X-rays\\Volumetric imaging\end{tabular} &
0.58--0.77 &
No &
\begin{tabular}[t]{@{}l@{}}High cost\\Specialised infras-\\tructure\\Staff demanding\\Demanding post-\\processing\end{tabular} &
\begin{tabular}[t]{@{}l@{}}Internal structures and volumetric study\\Thickness measurements\\Acoustic modelling (in case of violins)\\Non-invasive measurements\\No surface texture\\Segmentation and post-processing\\Less competitive for surface geometry\end{tabular} \\
\addlinespace[6pt]

LiDAR (Scaniverse) &
\begin{tabular}[t]{@{}l@{}}Consumer LiDAR\\Time-of-flight\end{tabular} &
0.99--1.21 &
No &
\begin{tabular}[t]{@{}l@{}}$\sim$\EUR{1k}\\Consumer iPad Pro\\ Real-time feedback\\Fast and accessible\end{tabular} &
\begin{tabular}[t]{@{}l@{}}Rapid low-cost acquisition\\Contextual documentation\\Limited geometric fidelity\\Education and outreach\end{tabular} \\
\addlinespace[6pt]

LiDAR (Heges) &
\begin{tabular}[t]{@{}l@{}}Consumer LiDAR\\Time-of-flight\end{tabular} &
1.72--1.82 &
No &
\begin{tabular}[t]{@{}l@{}}$\sim$\EUR{1k}\\Consumer iPad Pro\\ Real-time feedback\\Fast and accessible\end{tabular} &
\begin{tabular}[t]{@{}l@{}}Rapid low-cost acquisition\\Contextual documentation\\Least accurate technique\\evaluated in this study\\Education and outreach\end{tabular} \\

\bottomrule
\end{tabular}

\caption{Summary of the main characteristics and recommended uses of the acquisition techniques evaluated in this study. The intrinsic-error ranges correspond to the global estimates obtained using the WNNLS and WNNLAD formulations (Tables~4 and~5) and should not be interpreted as absolute measurement accuracies.}

\label{tab:techniques_summary}

\vspace{1mm}
\begin{minipage}{0.98\textwidth}
\scriptsize
$^{a}$The global optimisation yields a null intrinsic error for $\mu$-CT in both weighted formulations. This value should not be interpreted as perfect accuracy, since $\mu$-CT is weakly represented in the dataset. Individual estimates reported in Tables~4 and~5 suggest an intrinsic error on the order of a few hundredths of a millimetre. 
\end{minipage}

\end{table*}

\section{Conclusion}

\noindent This study presented a comprehensive comparison of ten three-dimensional acquisition techniques, including several operational modes, applied to the digitisation of historical instruments from the violin family. By analysing 254 pairwise comparisons across multiple instruments, acquisition contexts, and reconstruction pipelines, we quantitatively and qualitatively assessed the intrinsic errors, practical constraints, and application domains associated with each technique. Rather than seeking to establish a universal ranking, the objective was to characterise the strengths and limitations of each method in realistic, non-ideal acquisition conditions, reflecting the diversity of workflows encountered in museum and research environments. 

\noindent Beyond the numerical results, this work highlights the inherently transdisciplinary nature of three-dimensional digitisation in cultural heritage. The acquisition and interpretation of 3D data lies at the intersection of technological sciences—such as computer vision, metrology, and signal processing—and human sciences, including organology, conservation, art history, and museology. The digitisation of musical instruments, in particular, requires balancing geometric accuracy, visual fidelity, interpretability, and practical feasibility, while respecting the physical and historical integrity of the objects. 

\noindent The results confirm that no single acquisition technique is universally optimal. Photogrammetry emerges as a robust and cost-effective solution for archival documentation and public dissemination, offering high geometric accuracy together with rich surface texture. Structured-light scanning provides a reliable alternative for precise surface digitisation with direct metric scaling, while $\mu$-CT remains indispensable for investigations requiring access to internal structures and volumetric information. Mobile LiDAR solutions, although less accurate, offer unprecedented accessibility and speed, making them well suited for education, outreach, and rapid surveys. In contrast, CT and TLS were found to be less competitive in the present context, due to their complexity, lack of texture information, or the significant post-processing effort required to obtain usable models. 

\noindent These findings are particularly relevant in the broader context of ongoing digital transformation initiatives in museums, many of which aim to digitise significant portions of their collections by 2030. High-quality 3D models are increasingly central not only to documentation and conservation strategies, but also to research, education, and public engagement, including virtual exhibitions and augmented reality applications. By providing a comparative, application-oriented framework, this study contributes to informed decision-making regarding the selection of acquisition technologies, helping institutions align their technical choices with their scientific, cultural, and societal objectives. 

\noindent Finally, while the present work focuses on violin-family instruments with specific geometric characteristics, the methodological approach and comparative perspective developed here can be extended to other classes of cultural heritage objects. Future work will aim to broaden the range of object types and acquisition scenarios, refine uncertainty quantification, and further integrate 3D digitisation workflows into interdisciplinary research and conservation practices.

\section*{Acknowledgements}

The authors would like to thank Iona Thys (then at Royal Museums of Art and History) and Julie Toupance (Musée de la Musique) for the first two photographic acquisitions described in this paper that led to many photogrammetric models. We also acknowledge Gwendal Deniel (then intern at Musée de la musique) and Julie Toupance for providing the blue-light structured-light models, Henry-Louis Guillaume (PANORAMA, Université Libre de Bruxelles) for the infrared-based models, Professors Emmanuel Coche and Etienne Danse (Saint-Luc University Hospital, UCLouvain) and their team, who volunteered their time to carry out the CT scans, and finally Marta Bellato and Patricia Wils (AST-RX platform, UAR 2047 DoHNEE CNRS-MNHN, Paris) for the $\mu$-CT scans. All acquisitions carried out at the Musical Instruments Museum in Brussels were made possible thanks to the ingenuity of Joris De Valck, who designed and implemented several custom devices to stabilise the instruments while allowing accurate and safe data acquisition. Finally, we thank Pauline Terlinden (UCLouvain) for the preliminary results of her master’s thesis that contributed to this study, and for the conversion of CT and $\mu$-CT scan slices into three-dimensional models.

\appendix 
\section{Point clouds size and point-to-plane distances}
\label{section:appendix}

Table \ref{tab:point cloud sizes} indicates the number of points in each point cloud used in this study while Tables \ref{tab:pairwise_distances_1} and \ref{tab:pairwise_distances_2} show the point-to-plane distances $\sqrt{D^2_{plane}}$ (in \SI{}{\mm}) computed for the smaller instruments (1355 and 1731), and the point-to-plane distances $\sqrt{D^2_{plane}}$ for the larger instruments (2020, 2863 and 1374). Tables \ref{tab:pairwise_distances_1} and \ref{tab:pairwise_distances_2} therefore summarise the 254 comparisons performed in the paper.

\begin{table*}
\centering
\begin{tabular}{lcccccccccc}
\toprule
\textbf{Technique} &
\textbf{1355 sb} &
\textbf{1355 back} &
\textbf{1731 sb} &
\textbf{1731 back} &
\textbf{2020 sb} &
\textbf{2020 back} &
\textbf{2863 sb} &
\textbf{2863 back} &
\textbf{1374 sb} &
\textbf{1374 back} \\
\midrule
$\mu$-CT
& -     & -
& \phantom{0}281239    & \phantom{0}328088
& -     & -
& -     & -
& -     & - \\

SL (BL)
& -     & -
& \phantom{0}240983     & \phantom{0}271279
& 2478999     & \phantom{0}209029
& -     & -
& \phantom{0}636726     & - \\

Photog. PAN.
& 1654807     & 1634254
& -     & -
& -     & -
& 1376600     & 1044251
& -     & - \\

SL (IR)
& 1347585     & 1583387
& -     & -
& -     & -
& \phantom{0}357918     & \phantom{0}637693
& 1631561     & - \\

Photog.
& \phantom{0}130985     & \phantom{0}146238
& \phantom{0}207284     & \phantom{0}151053
& \phantom{0}106229     & 1267803
& -     & -
& \phantom{0}156306     & \phantom{0}151053 \\

TrueDepth
& \phantom{00}66750     & \phantom{00}68330
& -     & -
& -     & -
& \phantom{0}257305     & \phantom{0}286677
& \phantom{0}303933     & \phantom{0}344548 \\

TLS
& \phantom{00}16651     & \phantom{00}37676
& -     & -
& -     & -
& \phantom{0}141281     & \phantom{0}154510
& -     & - \\

CT
& \phantom{0}235665     & \phantom{0}250906
& -     & -
& -     & -
& \phantom{0}636777     & \phantom{0}679942
& -     & - \\

LiDAR (S)
& \phantom{000}5854     & \phantom{00}10623
& -     & -
& -     & -
& \phantom{00}19314     & \phantom{00}14367
& \phantom{00}18914     & \phantom{000}5515 \\

LiDAR (H)
& \phantom{00}65771     & \phantom{00}75199
& -     & -
& -     & -
& \phantom{0}311630     & \phantom{0}334095
& \phantom{0}377338     & \phantom{0}432459 \\
\bottomrule
\end{tabular}
\caption{Number of points in the point clouds used in this study. Each instrument is referred by its inventory number and is separated for each plate ('sb' refers to the sound board, while 'back' explicitly indicates the back)}
\label{tab:point cloud sizes}
\end{table*}

\begin{table*}[htbp]
\centering
\setlength{\tabcolsep}{4pt}
\renewcommand{\arraystretch}{1.15}
\small
\begin{tabular}{cllcccc}
\toprule
\textbf{Instrument} & \textbf{Technology 1} & \textbf{Technology 2} & \multicolumn{2}{c}{\textbf{Sound board}} & \multicolumn{2}{c}{\textbf{Back}} \\
\cmidrule(lr){4-5}\cmidrule(lr){6-7}
 &  &  & $d(\mathrm{tech}_1,\mathrm{tech}_2)$ & $d(\mathrm{tech}_2,\mathrm{tech}_1)$ & $d(\mathrm{tech}_1,\mathrm{tech}_2)$ & $d(\mathrm{tech}_2,\mathrm{tech}_1)$ \\
\midrule

\multirow{28}{*}{\textbf{1355}} 
 & Photogrammetry & Photogrammetry PANORAMA & 0.17495 & 0.13671 & 0.15958 & 0.09341 \\
 & Photogrammetry & CT-scan & 0.85766 & 0.60224 & 0.83469 & 0.78087 \\
 & Photogrammetry & Structured-light (IR) & 0.19277 & 0.07887 & 0.14804 & 0.07473 \\
 & Photogrammetry & TLS & 0.73968 & 0.86855 & 0.52522 & 0.76083 \\
 & Photogrammetry & LiDAR Scaniverse & 1.33309 & 1.03476 & 0.60493 & 0.58532 \\
 & Photogrammetry & LiDAR Heges & 4.03357 & 1.54665 & 2.90394 & 1.27910 \\
 & Photogrammetry & TrueDepth & 0.31999 & 0.20691 & 0.27575 & 0.20716 \\
 & Photogrammetry PANORAMA & CT-scan & 0.53673 & 0.49568 & 0.65879 & 0.74079 \\
 & Photogrammetry PANORAMA & Structured-light (IR) & 0.13561 & 0.04352 & 0.04988 & 0.04041 \\
 & Photogrammetry PANORAMA & TLS & 0.87069 & 0.83852 & 0.46963 & 0.73417 \\
 & Photogrammetry PANORAMA & LiDAR Scaniverse & 1.18433 & 1.03008 & 0.58202 & 0.59425 \\
 & Photogrammetry PANORAMA & LiDAR Heges & 1.81734 & 1.45909 & 1.43535 & 1.25192 \\
 & Photogrammetry PANORAMA & TrueDepth & 0.26576 & 0.18769 & 0.23178 & 0.21647 \\
 & CT-scan & Structured-light (IR) & 0.54898 & 0.50163 & 0.75495 & 0.64164 \\
 & CT-scan & TLS & 1.02837 & 0.97150 & 0.56936 & 0.90649 \\
 & CT-scan & LiDAR Scaniverse & 1.29712 & 1.12716 & 0.97235 & 0.97303 \\
 & CT-scan & LiDAR Heges & 1.97439 & 1.46700 & 1.78548 & 1.53719 \\
 & CT-scan & TrueDepth & 0.59800 & 0.52322 & 0.60866 & 0.48910 \\
 & Structured-light (IR) & TLS & 0.66592 & 0.86219 & 0.46695 & 0.73530 \\
 & Structured-light (IR) & LiDAR Scaniverse & 0.99649 & 1.04159 & 0.55589 & 0.63002 \\
 & Structured-light (IR) & LiDAR Heges & 1.51870 & 1.53716 & 1.43434 & 1.26305 \\
 & Structured-light (IR) & TrueDepth & 0.18733 & 0.21107 & 0.22312 & 0.20733 \\ 
 & TLS & LiDAR Scaniverse & 1.34047 & 1.08771 & 0.91382 & 0.67286 \\
 & TLS & LiDAR Heges & 2.14772 & 1.64724 & 1.99474 & 1.43836 \\
 & TLS & TrueDepth & 0.84831 & 0.63888 & 0.69447 & 0.45767 \\ 
 & LiDAR Scaniverse & LiDAR Heges & 2.33649 & 1.61398 & 2.17905 & 1.43388 \\
 & LiDAR Scaniverse & TrueDepth & 1.12392 & 0.95701 & 0.65136 & 0.59497 \\ 
 & LiDAR Heges & TrueDepth & 1.67180 & 1.61375 & 1.47182 & 1.76978 \\
\addlinespace[2pt]

\midrule

\multirow{3}{*}{\textbf{1731}} & Photogrammetry & $\mu$CT-scan & 0.34598 & 0.08419 & 0.12114 & 0.11690 \\
 & Photogrammetry & Structured-light (BL) & 0.17612 & 0.08804 & 0.12808 & 0.09275 \\
 & $\mu$CT-scan & Structured-light (BL) & 0.06976 & 0.08649 & 0.11525 & 0.10260 \\
\addlinespace[2pt]

\bottomrule
\end{tabular}
\caption{Point-to-plane distances $\sqrt{D^2_{plane}}$ between point clouds of the smaller instrument from different acquisition techniques (in \SI{}{\mm}).}
\label{tab:pairwise_distances_1}
\end{table*}

\begin{table*}[htbp]
\centering
\setlength{\tabcolsep}{4pt}
\renewcommand{\arraystretch}{1.15}
\small
\begin{tabular}{cllcccc}
\toprule
\textbf{Instrument} & \textbf{Technology 1} & \textbf{Technology 2} & \multicolumn{2}{c}{\textbf{Sound board}} & \multicolumn{2}{c}{\textbf{Back}} \\
\cmidrule(lr){4-5}\cmidrule(lr){6-7}
 &  &  & $d(\mathrm{tech}_1,\mathrm{tech}_2)$ & $d(\mathrm{tech}_2,\mathrm{tech}_1)$ & $d(\mathrm{tech}_1,\mathrm{tech}_2)$ & $d(\mathrm{tech}_2,\mathrm{tech}_1)$ \\
\midrule

\multirow{1}{*}{\textbf{2020}} 
 & Photogrammetry & Structured-light (BL) 
 & 0.48811 & 0.13907 
 & 0.37927 & 0.30502 \\
\addlinespace[2pt]

\midrule
 
\multirow{21}{*}{\textbf{2863}} 
 & Photogrammetry PANORAMA & CT-scan & 0.99486 & 1.14594 & 1.12745 & 1.70536 \\
 & Photogrammetry PANORAMA & Structured-light (IR) & 0.22736 & 0.32025 & 0.34648 & 0.53445 \\
 & Photogrammetry PANORAMA & TLS & 0.63184 & 0.72666 & 0.75643 & 0.65875 \\
 & Photogrammetry PANORAMA & LiDAR Scaniverse & 2.09137 & 1.36982 & 1.45875 & 1.62484 \\
 & Photogrammetry PANORAMA & LiDAR Heges & 1.67400 & 1.87115 & 2.75869 & 1.71369 \\
 & Photogrammetry PANORAMA & TrueDepth & 1.03215 & 0.87552 & 1.21815 & 0.89629 \\
 & CT-scan & Structured-light (IR) & 1.16012 & 1.00888 & 1.59257 & 2.10340 \\
 & CT-scan & TLS & 1.20224 & 1.17180 & 1.38488 & 1.20778 \\
 & CT-scan & LiDAR Scaniverse & 2.07498 & 1.56569 & 2.27448 & 1.74169 \\
 & CT-scan & LiDAR Heges & 2.03259 & 1.98109 & 2.73741 & 1.95022 \\
 & CT-scan & TrueDepth & 1.43365 & 1.34569 & 1.45684 & 0.95656 \\
 & Structured-light (IR) & TLS & 0.92433 & 0.77836 & 0.97744 & 0.71987 \\
 & Structured-light (IR) & LiDAR Scaniverse & 1.85690 & 1.43760 & 1.59661 & 1.72368 \\
 & Structured-light (IR) & LiDAR Heges & 2.31287 & 1.97249 & 2.93203 & 1.71866 \\
 & Structured-light (IR) & TrueDepth & 1.27758 & 0.84298 & 1.50830 & 0.94588 \\
 & TLS & LiDAR Scaniverse & 2.21959 & 1.32987 & 1.49729 & 1.83577 \\
 & TLS & LiDAR Heges & 1.65406 & 1.81469 & 1.99962 & 1.41248 \\
 & TLS & TrueDepth & 1.13379 & 0.74761 & 1.15767 & 0.82877 \\
 & LiDAR Scaniverse & LiDAR Heges & 3.14089 & 2.30268 & 4.14876 & 1.99879 \\
 & LiDAR Scaniverse & TrueDepth & 1.51959 & 1.74905 & 1.83626 & 1.52003 \\
 & LiDAR Heges & TrueDepth & 2.41506 & 1.51037 & 2.56105 & 2.31754 \\
\addlinespace[2pt]

\midrule 

\multirow{15}{*}{\textbf{1374}} 
 & Photogrammetry & Structured-light (BL) & 0.41300 & 0.31540  & \cellcolor{lightgray} & \cellcolor{lightgray} \\
 & Photogrammetry & Structured-light (IR) & 0.85086 & 0.66516 & \cellcolor{lightgray} & \cellcolor{lightgray} \\
 & Photogrammetry & LiDAR Scaniverse & 1.76735 & 1.17265  & 1.50715 & 2.12348 \\
 & Photogrammetry & LiDAR Heges & 6.49885 & 2.08938  & 4.06473 & 2.00851 \\
 & Photogrammetry & TrueDepth & 1.04135 & 0.65497  & 1.07723 & 0.91951 \\
 & Structured-light (BL) & Structured-light (IR) & 0.51525 & 0.32885  & \cellcolor{lightgray} & \cellcolor{lightgray} \\
 & Structured-light (BL) & LiDAR Scaniverse & 1.43185 & 1.21324  & \cellcolor{lightgray} & \cellcolor{lightgray} \\
 & Structured-light (BL) & LiDAR Heges & 2.43424 & 1.88461  & \cellcolor{lightgray} & \cellcolor{lightgray} \\
 & Structured-light (BL) & TrueDepth & 0.54690 & 0.48704  & \cellcolor{lightgray} & \cellcolor{lightgray} \\
 & Structured-light (IR) & LiDAR Scaniverse & 1.61549 & 1.46779  & \cellcolor{lightgray} & \cellcolor{lightgray} \\
 & Structured-light (IR) & LiDAR Heges & 3.63088 & 1.96502 & \cellcolor{lightgray} & \cellcolor{lightgray} \\
 & Structured-light (IR) & TrueDepth & 0.53195 & 0.53311 & \cellcolor{lightgray} & \cellcolor{lightgray} \\
 & LiDAR Scaniverse & LiDAR Heges & 4.91046 & 2.08768 & 3.83578 & 1.83830 \\
 & LiDAR Scaniverse & TrueDepth & 1.13034 & 1.11098 & 2.36237 & 1.42207 \\
 & LiDAR Heges & TrueDepth & 1.67711 & 2.05872 & 2.01273 & 2.09592 \\
\addlinespace[2pt]

\bottomrule
\end{tabular}
\caption{Point-to-plane distances $\sqrt{D^2_{plane}}$ between point clouds of the larger instruments from different acquisition techniques (in \SI{}{\mm}).}
\label{tab:pairwise_distances_2}
\end{table*}

\bibliographystyle{eg-alpha-doi} 
\bibliography{egbibsample}       

\newcommand{\etalchar}[1]{$^{#1}$}
\begin{thebibliography}{\uppercase{MBPRLC{\etalchar{*}}21}}

\bibitem[ACLN18]{aicardi2018recent}
\textsc{Aicardi I., Chiabrando F., Lingua A.~M., Noardo F.}:
\newblock Recent trends in cultural heritage {3D} survey: The photogrammetric
  computer vision approach.
\newblock \emph{Journal of Cultural Heritage 32} (2018), 257--266.

\bibitem[AMSAWA24]{abdel2024indoor}
\textsc{Abdel-Majeed H.~M., Shaker I.~F., Abdel-Wahab A., Awad A. A. D.~I.}:
\newblock Indoor mapping accuracy comparison between the {A}pple devices’
  {L}i{DAR} sensor and terrestrial laser scanner.
\newblock \emph{HBRC Journal 20}, 1 (2024), 915--931.

\bibitem[ARA20]{adamopoulos2020critical}
\textsc{Adamopoulos E., Rinaudo F., Ardissono L.}:
\newblock A critical comparison of 3{D} digitization techniques for heritage
  objects.
\newblock \emph{ISPRS International Journal of Geo-Information 10}, 1 (2020),
  10.

\bibitem[B{\'A}F{\etalchar{*}}25]{beltran2025performance}
\textsc{Beltr{\'a}n N., {\'A}lvarez J., Fern{\'a}ndez A., {\'A}lvarez B.~J.,
  Blanco D.}:
\newblock Performance analysis of {A}pple’s {T}rue{D}epth sensor for surface
  reconstruction of regular geometries across varying stand-off distances.
\newblock \emph{The International Journal of Advanced Manufacturing Technology}
  (2025), 1--9.

\bibitem[BCFG23]{beghin2023validation}
\textsc{Beghin P., Ceulemans A.-E., Fisette P., Glineur F.}:
\newblock Validation of a photogrammetric approach for the objective study of
  early bowed instruments.
\newblock \emph{Heritage Science 11}, 1 (2023), 170.

\bibitem[BCG25]{beghin2025identification}
\textsc{Beghin P., Ceulemans A.-E., Glineur F.}:
\newblock Identification of violin reduction via contour lines classification.
\newblock \emph{arXiv preprint arXiv:2507.07743} (2025).

\bibitem[BCG26]{beghin2026discussion}
\textsc{Beghin P., Ceulemans A.-E., Glineur F.}:
\newblock A discussion about violin reduction: Geometric analysis of contour
  lines and channel of minima.
\newblock In \emph{The Future of Heritage Science and Technologies II} (Cham,
  2026), Furferi R., Governi L., Volpe Y., Gherardini F., Seymour K., (Eds.),
  Springer Nature Switzerland, pp.~432--446.

\bibitem[Beg21]{beghindigital}
\textsc{Beghin P.}:
\newblock \emph{A digital tool at the service of organology: validation of a
  photogrammetric approach}.
\newblock Master's thesis, Ecole polytechnique de Louvain, UCLouvain, 2021.

\bibitem[BM92]{besl1992method}
\textsc{Besl P.~J., McKay N.~D.}:
\newblock Method for registration of 3-{D} shapes.
\newblock In \emph{Sensor fusion IV: control paradigms and data structures}
  (1992), vol.~1611, Spie, pp.~586--606.

\bibitem[Bro22]{brown2022old}
\textsc{Brown M.-E.}:
\newblock Old meets new: 3-{D} printing and the art of violin-making.
\newblock \emph{The Journal of the Acoustical Society of America 152},
  4\_Supplement (2022), A93--A93.

\bibitem[BS09]{borman2009review}
\textsc{Borman T., Stoel B.}:
\newblock Review of the uses of computed tomography for analyzing instruments
  of the violin family with a focus on the future.
\newblock \emph{J Violin Soc Am: VSA Papers 22}, 1 (2009), 1--12.

\bibitem[CBF{\etalchar{*}}23]{ceulemans2023baroque}
\textsc{Ceulemans A.-E., Beghin P., Fisette P., Glineur F., Thys I.}:
\newblock Baroque violas with reduced soundboxes: An evaluation method.
\newblock \emph{The Galpin Society Journal 76 (LXXVI)} (2023), 109--126.

\bibitem[CDR15]{chiabrando2015sfm}
\textsc{Chiabrando F., Donadio E., Rinaudo F.}:
\newblock {S}f{M} for orthophoto to generation: {A} winning approach for
  cultural heritage knowledge.
\newblock \emph{The international archives of the photogrammetry, remote
  sensing and spatial information sciences 40} (2015), 91--98.

\bibitem[CM92]{chen1992object}
\textsc{Chen Y., Medioni G.}:
\newblock Object modelling by registration of multiple range images.
\newblock \emph{Image and vision computing 10}, 3 (1992), 145--155.

\bibitem[CP23]{cavalieri2023so}
\textsc{Cavalieri M., Peeters A.}:
\newblock The so-called insula dei {L}ottatori and block {V}, {III} in {O}stia:
  the combination of terrestrial laser scanner ({TLS}) and drones ({UAS}) in
  the field of cultural heritage.
\newblock \emph{GROMA: Documenting Archaeology 8} (2023), 110--127.

\bibitem[CS13]{chiabrando2013points}
\textsc{Chiabrando F., Span{\`o} A.}:
\newblock Points clouds generation using {TLS} and dense-matching techniques.
  {A} test on approachable accuracies of different tools.
\newblock \emph{ISPRS annals of the photogrammetry, remote sensing and spatial
  information sciences 2} (2013), 67--72.

\bibitem[DHA{\etalchar{*}}18]{daneshmand20183d}
\textsc{Daneshmand M., Helmi A., Avots E., Noroozi F., Alisinanoglu F., Arslan
  H.~S., Gorbova J., Haamer R.~E., Ozcinar C., Anbarjafari G.}:
\newblock {3D} scanning: A comprehensive survey.
\newblock \emph{arXiv preprint arXiv:1801.08863} (2018).

\bibitem[DLML16]{dondi2016measuring}
\textsc{Dondi P., Lombardi L., Malagodi M., Licchelli M.}:
\newblock Measuring {S}tradivari ‘{C}remonese’(1715) by {3D} modeling.
\newblock In \emph{IMEKO International Conference on Metrology for Archeology
  and Cultural Heritage} (2016), MetroArcheo, pp.~29--33.

\bibitem[DLML17]{dondi20173d}
\textsc{Dondi P., Lombardi L., Malagodi M., Licchelli M.}:
\newblock {3D} modelling and measurements of historical violins.
\newblock \emph{Acta IMEKO 6}, 3 (2017), 29--34.

\bibitem[DLR{\etalchar{*}}18]{dondi2018multimodal}
\textsc{Dondi P., Lombardi L., Rocca I., Malagodi M., Licchelli M.}:
\newblock Multimodal workflow for the creation of interactive presentations of
  360 spin images of historical violins.
\newblock \emph{Multimedia Tools and Applications 77}, 21 (2018), 28309--28332.

\bibitem[Dra12]{drap2012underwater}
\textsc{Drap P.}:
\newblock Underwater photogrammetry for archaeology.
\newblock \emph{Special applications of photogrammetry 114} (2012).

\bibitem[DSF{\etalchar{*}}22]{dollins2022design}
\textsc{Dollins C.~M., Scruton M., Frischling E. R.~B., O'Grady J.~P., Sullivan
  J.~M.}:
\newblock Design, analysis, and fabrication of a 3{D} printed violin for the
  public.
\newblock In \emph{ASEE-NE 2022} (2022).

\bibitem[{\'E}ch20a]{echard2020bass_violin}
\textsc{{\'E}chard J.-P.}:
\newblock Basse de violon recoupée en violoncelle.
\newblock In \emph{Stradivarius et la lutherie de Crémone}. Philharmonie de
  Paris, 2020, pp.~26--33.

\bibitem[{\'E}ch20b]{echard2020tenor_violin}
\textsc{{\'E}chard J.-P.}:
\newblock Ténor de violon recoupé en alto.
\newblock In \emph{Stradivarius et la lutherie de Crémone}. Philharmonie de
  Paris, 2020, pp.~20--25.

\bibitem[FBKC{\etalchar{*}}12]{fedorov20123d}
\textsc{Fedorov A., Beichel R., Kalpathy-Cramer J., Finet J., Fillion-Robin
  J.-C., Pujol S., Bauer C., Jennings D., Fennessy F., Sonka M., et~al.}:
\newblock {3D} {S}licer as an image computing platform for the quantitative
  imaging network.
\newblock \emph{Magnetic resonance imaging 30}, 9 (2012), 1323--1341.

\bibitem[Fer24]{fernie20243d}
\textsc{Fernie K.}:
\newblock {3D} content in {E}uropeana: {T}he challenges of providing access.
\newblock In \emph{The 3 Dimensions of Digitalised Archaeology:
  State-of-the-Art, Data Management and Current Challenges in Archaeological
  3D-Documentation}. Springer, 2024, pp.~167--177.

\bibitem[FGC12]{fioravanti2012structural}
\textsc{Fioravanti M., Goli G., Carlson B.}:
\newblock Structural assessment and measurement of the elastic deformation of
  historical violins: the case study of the {G}uarneri ‘del {G}es{\`u}’
  violin (1743) known as the ‘{C}annone’.
\newblock \emph{Journal of cultural heritage 13}, 2 (2012), 145--153.

\bibitem[FGI{\etalchar{*}}21]{fiocco2021compositional}
\textsc{Fiocco G., Gonzalez S., Invernizzi C., Rovetta T., Albano M., Dondi P.,
  Licchelli M., Antonacci F., Malagodi M.}:
\newblock Compositional and morphological comparison among three coeval violins
  made by {G}iuseppe {G}uarneri “del {G}es{\`u}” in 1734.
\newblock \emph{Coatings 11}, 8 (2021), 884.

\bibitem[FMC{\etalchar{*}}22]{fiz2022examples}
\textsc{Fiz J.~I., Mart{\'\i}n P.~M., Cuesta R., Sub{\'\i}as E., Codina D.,
  Cartes A.}:
\newblock Examples and results of aerial photogrammetry in archeology with
  {UAV}: Geometric documentation, high resolution multispectral analysis,
  models and {3D} printing.
\newblock \emph{Drones 6}, 3 (2022), 59.

\bibitem[FMR{\etalchar{*}}22]{farella2022analysing}
\textsc{Farella E.~M., Morelli L., Rigon S., Grilli E., Remondino F.}:
\newblock Analysing key steps of the photogrammetric pipeline for museum
  artefacts {3D} digitisation.
\newblock \emph{Sustainability 14}, 9 (2022), 5740.

\bibitem[FMS{\etalchar{*}}23]{frost2023digital}
\textsc{Frost A., Mirashrafi S., S{\'a}nchez C.~M., Vacas-Madrid D., Millan
  E.~R., Wilson L.}:
\newblock Digital documentation of reflective objects: a cross-polarised
  photogrammetry workflow for complex materials.
\newblock In \emph{{3D} research challenges in cultural heritage III:
  Complexity and quality in digitisation}. Springer, 2023, pp.~131--155.

\bibitem[FSHF09]{frohlich2009secrets}
\textsc{Frohlich B., Sturm G., Hinton J., Frohlich E.}:
\newblock The secrets of the {S}tradivari string instruments. a non-destructive
  study of music instruments from the {S}mithsonian {I}nstitution, the
  {L}ibrary of congress, and private collections. {A} pilot study of seven
  violins made by {A}ntonio {S}tradivari in {C}remona, {I}taly, between 1677
  and 1709.
\newblock \emph{Ann Arbor, MI, USA, Washington, DC (March 2009)} (2009).

\bibitem[FWK{\etalchar{*}}18]{fuchs2018musices}
\textsc{Fuchs T., Wagner R., Kretzer C., Scholz G., B{\"a}r F., Kirsch S.,
  Wolters-Rosbach M., Fischeidl K.}:
\newblock Musices-musical instrument computed tomography examination standard:
  Results of the measurements and guidelines derived therefrom.
\newblock \emph{Gothenburg, Schweden} (2018).

\bibitem[GALS17]{graham2017epic}
\textsc{Graham C.~A., Akoglu K.~G., Lassen A.~W., Simon S.}:
\newblock Epic dimensions: a comparative analysis of 3{D} acquisition methods.
\newblock \emph{The International Archives of the Photogrammetry, Remote
  Sensing and Spatial Information Sciences 42} (2017), 287--293.

\bibitem[GBA04]{guidi2004high}
\textsc{Guidi G., Beraldin J.-A., Atzeni C.}:
\newblock High-accuracy {3D} modeling of cultural heritage: the digitizing of
  {D}onatello's "{M}addalena".
\newblock \emph{IEEE Transactions on image processing 13}, 3 (2004), 370--380.

\bibitem[GGH{\etalchar{*}}21]{gebler2021comparison}
\textsc{Gebler O.~F., Goudswaard M., Hicks B., Jones D., Nassehi A., Snider C.,
  Yon J.}:
\newblock A comparison of structured light scanning and photogrammetry for the
  digitisation of physical prototypes.
\newblock \emph{Proceedings of the Design Society 1} (2021), 11--20.

\bibitem[GPBR15]{glira2015a}
\textsc{Glira P., Pfeifer N., Briese C., Ressl C.}:
\newblock A correspondence framework for {ALS} strip adjustments based on
  variants of the {ICP} algorithm.
\newblock \emph{Photogrammetrie-Fernerkundung-Geoinformation 2015}, 4 (2015),
  275--289.

\bibitem[GSAS21]{gonzalez2021eigenfrequency}
\textsc{Gonzalez S., Salvi D., Antonacci F., Sarti A.}:
\newblock Eigenfrequency optimisation of free violin plates.
\newblock \emph{The Journal of the Acoustical Society of America 149}, 3
  (2021), 1400--1410.

\bibitem[HG02]{hanke2002architectural}
\textsc{Hanke K., Grussenmeyer P.}:
\newblock Architectural photogrammetry: basic theory, procedures, tools.
\newblock In \emph{ISPRS Commission} (2002), vol.~5, Taylor \& Francis Abington
  UK, pp.~1--2.

\bibitem[HR10]{hess20103d}
\textsc{Hess M., Robson S.}:
\newblock {3D} colour imaging for cultural heritage artefacts.
\newblock \emph{International Archives of Photogrammetry, Remote Sensing and
  Spatial Information Sciences 38}, Part 5 (2010), 288--292.

\bibitem[IFB{\etalchar{*}}24]{isaac2024making}
\textsc{Isaac A., Fernie K., Bachi V., Tsoupra E., Medici M., Alkemade H.,
  M{\"u}nster S., Charles V., Heslinga L.}:
\newblock Making the {E}uropeana data model a better fit for documentation of
  {3D} objects.
\newblock In \emph{3D Research Challenges in Cultural Heritage V: Paradata,
  Metadata and Data in Digitisation}. Springer, 2024, pp.~63--74.

\bibitem[KAMK18]{khalaf20183d}
\textsc{Khalaf A., Ataiwe T., Mohammed I., Kareem A.}:
\newblock {3D} digital modeling for archeology using close range
  photogrammetry.
\newblock In \emph{MATEC Web of Conferences} (2018), vol.~162, EDP Sciences,
  p.~03027.

\bibitem[Kin20]{kingsland2020comparative}
\textsc{Kingsland K.}:
\newblock Comparative analysis of digital photogrammetry software for cultural
  heritage.
\newblock \emph{Digital Applications in Archaeology and Cultural Heritage 18}
  (2020), e00157.

\bibitem[Kir24]{kirsch2024sarasate}
\textsc{Kirsch S.}:
\newblock \emph{Technical drawing: Violin ``Sarasate''}.
\newblock Éditions de la Philharmonie de Paris, 2024.
\newblock URL:
  \url{https://librairie.philharmoniedeparis.fr/dessins-techniques/violon-sarasate}.

\bibitem[KM18]{kirsch2018x}
\textsc{Kirsch S., Mannes D.}:
\newblock X-ray {CT} and neutron imaging for musical instruments--a comparative
  study.
\newblock \emph{Wooden musical instruments - Different forms of knowledge -
  Book of end of woodMusICK - COST action FP1302} (2018), 157--170.

\bibitem[Kno87]{knott1987modal}
\textsc{Knott G.~A.}:
\newblock \emph{A modal analysis of the violin using MSC/NASTRAN and PATRAN}.
\newblock PhD thesis, Monterey, California: Naval Postgraduate School, 1987.

\bibitem[LCB{\etalchar{*}}25]{laurent2025combining}
\textsc{Laurent A., Coupry B., Brument B., M{\'e}lou J., Qu{\'e}au Y., Fritz
  C., Durou J.-D.}:
\newblock Combining geometric and photometric {3D} reconstruction techniques
  for cultural heritage.
\newblock \emph{Journal of Cultural Heritage 73} (2025), 43--51.

\bibitem[Lot19]{lothairecharacterization}
\textsc{Lothaire R.}:
\newblock \emph{Characterization of violins: a digital tool at the service of
  organology}.
\newblock Master's thesis, Ecole polytechnique de Louvain, UCLouvain, 2019.

\bibitem[MBPRLC{\etalchar{*}}21]{marin2021photogrammetry}
\textsc{Mar{\'\i}n-Buz{\'o}n C., P{\'e}rez-Romero A., L{\'o}pez-Castro J.~L.,
  Ben~Jerbania I., Manzano-Agugliaro F.}:
\newblock Photogrammetry as a new scientific tool in archaeology: Worldwide
  research trends.
\newblock \emph{Sustainability 13}, 9 (2021), 5319.

\bibitem[MBS13]{mathys2013comparing}
\textsc{Mathys A., Brecko J., Semal P.}:
\newblock Comparing {3D} digitizing technologies: what are the differences?
\newblock In \emph{2013 Digital Heritage International Congress
  (DigitalHeritage)} (2013), vol.~1, IEEE, pp.~201--204.

\bibitem[MDS{\etalchar{*}}20]{magnani2020digital}
\textsc{Magnani M., Douglass M., Schroder W., Reeves J., Braun D.~R.}:
\newblock The digital revolution to come: Photogrammetry in archaeological
  practice.
\newblock \emph{American Antiquity 85}, 4 (2020), 737--760.

\bibitem[MNR{\etalchar{*}}16]{menna20163d}
\textsc{Menna F., Nocerino E., Remondino F., Dellepiane M., Callieri M.,
  Scopigno R.}:
\newblock {3D} digitization of an heritage masterpiece-a critical analysis on
  quality assessment.
\newblock \emph{The International Archives of the Photogrammetry, Remote
  Sensing and Spatial Information Sciences 41} (2016), 675--683.

\bibitem[Mot17]{motteconception}
\textsc{Motte L.}:
\newblock \emph{Conception d’un outil math\'ematique/informatique pour la
  classification d’instruments \`a archet}.
\newblock Master's thesis, Ecole polytechnique de Louvain, UCLouvain, 2017.

\bibitem[MPSV{\etalchar{*}}25]{meza2025towards}
\textsc{Meza-P{\'e}rez A., Stoppani G., Viala R., Gonzalez S., Igartua U., Rico
  R.~J., Fritz C.}:
\newblock Towards a finite element model of a batch of experimental violins:
  validation on sub-structural components.
\newblock \emph{Acta Acustica 9} (2025), 2.

\bibitem[MRW18]{marschke2018modeling}
\textsc{Marschke S., Ring W., Wohlmuth B.}:
\newblock Modeling, identification, and optimization of violin bridges.
\newblock \emph{PAMM 18}, 1 (2018), e201800405.

\bibitem[MWR{\etalchar{*}}20]{marschke2020approach}
\textsc{Marschke S., Wunderlich L., Ring W., Achterhold K., Pfeiffer F.}:
\newblock An approach to construct a three-dimensional isogeometric model from
  $\mu$-{CT} scan data with an application to the bridge of a violin.
\newblock \emph{Computer Aided Geometric Design 78} (2020), 101815.

\bibitem[PCDM18]{patrucco2018image}
\textsc{Patrucco G., Chiabrando F., Dondi P., Malagodi M.}:
\newblock Image and range-based {3D} acquisition and modeling of popular
  musical instruments.
\newblock \emph{Proceedings from the Document Academy 5}, 2 (2018), 9.

\bibitem[PCF19]{polo2019study}
\textsc{Polo M.-E., Cuartero A., Felic{\'\i}simo {\'A}.~M.}:
\newblock Study of uncertainty and repeatability in structured-light {3D}
  scanners.
\newblock \emph{arXiv preprint arXiv:1910.13199} (2019).

\bibitem[Pia17]{piantadosi2017three}
\textsc{Piantadosi S.}:
\newblock Three dimensional mathematical modeling of violin plate surfaces: An
  approach based on an ensemble of contour lines.
\newblock \emph{{PLoS} {O}ne 12}, 2 (2017), e0171167.

\bibitem[PK17]{plath2017post}
\textsc{Plath N., Kirsch S.}:
\newblock Post-processing of musical instrument {3D}-computed tomography data
  for conservational applications.
\newblock In \emph{Proc. WoodMusICK FP 1302 Cost Action Conf} (2017).

\bibitem[PKB11]{pyrkosz2011converting}
\textsc{Pyrkosz M., Karsen C.~V., Bissinger G.}:
\newblock Converting {CT} scans of a {S}tradivari violin to a {FEM}.
\newblock In \emph{Structural Dynamics, Volume 3: Proceedings of the 28th IMAC,
  A Conference on Structural Dynamics, 2010} (2011), Springer, pp.~811--820.

\bibitem[Pla19]{plath20193d}
\textsc{Plath N.}:
\newblock {3D} imaging of musical instruments: Methods and applications.
\newblock \emph{Computational Phonogram Archiving} (2019), 321--334.

\bibitem[Ple12]{pletinckx2012europeana}
\textsc{Pletinckx D.}:
\newblock {E}uropeana and {3D}.
\newblock \emph{The International Archives of the Photogrammetry, Remote
  Sensing and Spatial Information Sciences 38} (2012), 483--490.

\bibitem[PRF08]{pinto2008photogrammetric}
\textsc{Pinto L., Roncella R., Forlani G.}:
\newblock Photogrammetric survey of ancients musical instruments.
\newblock In \emph{The International Archives of the Photogrammetry, Remote
  Sensing and Spatial Information Sciences, ISPRS Congress, vol. XXXVII, Part
  B5} (2008), pp.~309--314.

\bibitem[PVK13]{pyrkosz2013comparative}
\textsc{Pyrkosz M., Van~Karsen C.}:
\newblock Comparative modal tests of a violin.
\newblock \emph{Experimental Techniques 37}, 4 (2013), 47--62.

\bibitem[PVK14]{pyrkosz2014coupled}
\textsc{Pyrkosz M.~A., Van~Karsen C.}:
\newblock Coupled vibro-acoustic model of the {T}itian {S}tradivari violin.
\newblock In \emph{Topics in Modal Analysis I, Volume 7: Proceedings of the
  32nd IMAC, A Conference and Exposition on Structural Dynamics, 2014} (2014),
  Springer, pp.~317--332.

\bibitem[RAMB{\etalchar{*}}24]{rocuts2024comparison}
\textsc{Rocuts A., Avella-Molano B., Behr A., Lakhani F., Bolds B.,
  Riveros-Amado M., Riveros-Perez E.}:
\newblock Comparison of two {3D} scanning software to identify facial features:
  a prospective instrument to predict difficult airway.
\newblock \emph{Perioperative Medicine 13}, 1 (2024), 9.

\bibitem[RDPC{\etalchar{*}}21]{roda20213d}
\textsc{Rod{\`a} A., De~Poli G., Canazza S., Sun Z., Whiting E., et~al.}:
\newblock {3D} virtual reconstruction and sound simulation of old musical
  instruments.
\newblock \emph{Archeologia e Calcolatori 32}, 1 (2021), 359--374.

\bibitem[Rem11]{remondino2011heritage}
\textsc{Remondino F.}:
\newblock Heritage recording and {3D} modeling with photogrammetry and {3D}
  scanning.
\newblock \emph{Remote sensing 3}, 6 (2011), 1104--1138.

\bibitem[Sap15]{sapirstein2015photogrammetry}
\textsc{Sapirstein P.}:
\newblock Photogrammetry as a tool for architectural analysis: the digital
  architecture project at {O}lympia.
\newblock In \emph{Archaeological Research in the Digital Age. Proceedings of
  the 1st Conference on Computer Applications and Quantitative Methods in
  Archaeology, Greek Chapter. Rethymno, Crete, March 6-8 2014} (2015),
  IMS-FORTH, Rethymno, pp.~129--139.

\bibitem[SB08]{stoel2008comparison}
\textsc{Stoel B.~C., Borman T.~M.}:
\newblock A comparison of wood density between classical {C}remonese and modern
  violins.
\newblock \emph{PLoS One 3}, 7 (2008), e2554.

\bibitem[SGC25]{schenkel2025photogrammetrie}
\textsc{Schenkel A., Guillaume H.-L., Clerbois S.}:
\newblock Photogramm{\'e}trie en polarisation crois{\'e}e: base d’analyse
  {3D} des processus d'industrialisation de la statuaire.
\newblock \emph{Patrimoines industriels immobiliers, mobiliers et
  immat{\'e}riels--Sauvetage et valorisation} (2025).

\bibitem[Sha98]{shakarji1998least}
\textsc{Shakarji C.~M.}:
\newblock Least-squares fitting algorithms of the {NIST} algorithm testing
  system.
\newblock \emph{Journal of research of the National Institute of Standards and
  Technology 103}, 6 (1998), 633.

\bibitem[SMD{\etalchar{*}}21]{stanciu2021x}
\textsc{Stanciu M.~D., Mih{\u{a}}lcic{\u{a}} M., Dinulic{\u{a}} F., Nauncef
  A.~M., Purdoiu R., L{\u{a}}c{\u{a}}tuș R., Gliga G.~V.}:
\newblock X-ray imaging and computed tomography for the identification of
  geometry and construction elements in the structure of old violins.
\newblock \emph{Materials 14}, 20 (2021), 5926.

\bibitem[SMG15]{schroeder2015flying}
\textsc{Schroeder W., Maynard R., Geveci B.}:
\newblock Flying edges: A high-performance scalable isocontouring algorithm.
\newblock In \emph{2015 IEEE 5th symposium on large data analysis and
  visualization (ldav)} (2015), IEEE, pp.~33--40.

\bibitem[SMG{\etalchar{*}}25]{schenkel2025preserving}
\textsc{Schenkel A., Muyle C., Guillaume H.-L., Reyniers J., Henderiks V.,
  Vercruysse L., Janssens K., Debeir O.}:
\newblock {Preserving the Sacred In Situ: A Scalable Model for Hybrid Religious
  Heritage Documentation}.
\newblock In \emph{Digital Heritage} (2025), Campana S., Ferdani D., Graf H.,
  Guidi G., Hegarty Z., Pescarin S., Remondino F., (Eds.), The Eurographics
  Association.
\newblock \href {https://doi.org/10.2312/dh.20253251}
  {\path{doi:10.2312/dh.20253251}}.

\bibitem[Ter25]{terlindencomparison}
\textsc{Terlinden P.}:
\newblock \emph{Comparison of 3{D} acquisition techniques for the digital
  reconstruction of bowed instruments}.
\newblock Master's thesis, Ecole polytechnique de Louvain, UCLouvain, 2025.

\bibitem[TGK{\etalchar{*}}21]{tsoraeva2021application}
\textsc{Tsoraeva E., Gadzhiev R., Kuchiev S., Pekh A., Mezhyan S.}:
\newblock Application of photogrammetric methods in architecture, construction
  and land management.
\newblock In \emph{IOP Conference Series: Materials Science and Engineering}
  (2021), vol.~1083, IOP Publishing.

\bibitem[TSTT20]{torres2020exploring}
\textsc{Torres J.~A., Soto C.~A., Torres-Torres D.}:
\newblock Exploring design variations of the {T}itian {S}tradivari violin using
  a finite element model.
\newblock \emph{The Journal of the Acoustical Society of America 148}, 3
  (2020), 1496--1506.

\bibitem[VRE21]{vogt2021comparison}
\textsc{Vogt M., Rips A., Emmelmann C.}:
\newblock Comparison of i{P}ad {P}ro’s {L}i{DAR} and {T}rue{D}epth
  capabilities with an industrial 3{D} scanning solution.
\newblock \emph{Technologies 9}, 2 (2021), 25.

\bibitem[YYGD07]{yilmaz2007importance}
\textsc{Yilmaz H.~M., Yakar M., Gulec S.~A., Dulgerler O.~N.}:
\newblock Importance of digital close-range photogrammetry in documentation of
  cultural heritage.
\newblock \emph{Journal of Cultural Heritage 8}, 4 (2007), 428--433.

\end{thebibliography}



\end{document}